\documentclass[a4paper,12pt]{article}
\usepackage{html,amsmath,amssymb,amsfonts,epsfig,cancel,cite}

\newlength{\xtrawidth}
\newlength{\xtraheight}
\newcommand{\be}{\begin{equation}}
\newcommand{\ee}{\end{equation}}
\newcommand{\beq}{\begin{equation}}
\newcommand{\eeq}{\end{equation}}
\newcommand{\ba}{\begin{array}}
\newcommand{\ea}{\end{array}}
\newcommand{\bea}{\begin{eqnarray}}
\newcommand{\eea}{\end{eqnarray}}
\newcommand{\bean}{\begin{eqnarray*}}
\newcommand{\eean}{\end{eqnarray*}}
\newcommand{\eref}[1]{(\ref{#1})}

\newcommand{\nn}{\nonumber}

\newcommand{\ch}{{\rm ch}}
\newcommand{\td}{{\rm Td}}
\newcommand{\ind}{\mathop{{\rm ind}}}

\newcommand{\IP}{\mathbb{P}}

\newcommand{\cO}{{\cal O}}

\newcommand{\cA}{{\cal A}}

\newcommand{\fn}{\footnotesize}

\newcommand{\anU}{{\cal J}}

\def\fnote#1#2{\begingroup\def\thefootnote{#1}\footnote{#2}
     \addtocounter{footnote}{-1}\endgroup}

\numberwithin{equation}{section}

\begin{document}

\vspace{1cm}

\title{{\Large \bf Heterotic Line Bundle Standard Models}}

\vspace{2cm}

\author{
Lara B. Anderson${}^{1}$,
James Gray${}^{2}$,
Andre Lukas${}^{3}$,
Eran Palti${}^{4}$
}
\date{}
\maketitle
\begin{center} {\small ${}^1${\it Center for the Fundamental Laws of Nature,  \\
Jefferson Laboratory, Harvard University, \\ 17 Oxford Street, Cambridge, MA 02138, U.S.A.}\\[0.2cm]
      ${}^2${\it Arnold-Sommerfeld-Center for Theoretical Physics, \\
       Department f\"ur Physik, Ludwig-Maximilians-Universit\"at M\"unchen,\\
       Theresienstra\ss e 37, 80333 M\"unchen, Germany}\\[0.2cm]
       ${}^3${\it Rudolf Peierls Centre for Theoretical Physics, Oxford
       University,\\
       $~~~~~$ 1 Keble Road, Oxford, OX1 3NP, U.K.\\[0.2cm]
       ${}^4${\it Centre de Physique Theorique, Ecole Polytechnique, CNRS, 91128 Palaiseau, France.}}}\\

\fnote{}{lara@physics.harvard.edu, james.gray@physik.uni-muenchen.de}
\fnote{}{lukas@physics.ox.ac.uk, Eran.Palti@cpht.polytechnique.fr} 

\thispagestyle{empty}
\setcounter{page}{0}

\end{center}

\abstract{\noindent }
In a previous publication, \htmladdnormallink{{\tt arXiv:1106.4804}}{http://arXiv.org/abs/arXiv:1106.4804}, we have found 200  models from heterotic Calabi-Yau compactifications with line bundles, which lead to standard models after taking appropriate quotients by a discrete symmetry and introducing Wilson lines. In this paper, we construct the resulting standard models explicitly, compute their spectrum including Higgs multiplets, and analyze some of their basic properties. After removing redundancies we find about 400 downstairs models, each with the precise matter spectrum of the supersymmetric standard model, with one, two or three pairs of Higgs doublets and no exotics of any kind. In addition to the standard model gauge group, up to four Green-Schwarz anomalous $U(1)$ symmetries are present in these models, which constrain the allowed operators in the four-dimensional effective supergravity. The vector bosons associated to these anomalous $U(1)$ symmetries are massive. We explicitly compute the spectrum of allowed operators for each model and present the results, together with the defining data of the models, in a database of standard models accessible \htmladdnormallink{{\tt here}}{http://www-thphys.physics.ox.ac.uk/projects/CalabiYau/linebundlemodels/index.html}. Based on these results we analyze elementary phenomenological properties. For example, for about 200 models all dimension four and five proton decay violating operators are forbidden by the additional $U(1)$ symmetries. 
\newpage

\tableofcontents

%

\section{Introduction}
Compactifications of the heterotic string on Calabi-Yau manifolds, despite being the oldest approach to string phenomenology~\cite{Candelas:1985en,gsw}, remains one of the most promising and well-understood paths to obtaining realistic string vacua. These models can combine the attractive ideas of grand unification with a large top Yukawa coupling, features which have proved to be difficult to realize in other types of models, particularly those based on type II string theory. In essence, this leaves the heterotic string, F-theory and the lesser studied $G_2$ compactifications of M-theory as primary starting points for string phenomenology.  

Traditionally, heterotic Calabi-Yau model building has been based on the standard embedding~\cite{Greene:1986bm,Greene:1986jb,Braun:2011ni} whereby the Bianchi identity is solved by setting the internal gauge bundle $V$ equal to the tangent bundle, $TX$, of the Calabi-Yau manifolds $X$. However, over the past decade it has been realized that this approach is too restrictive and the focus has shifted to the wider class of non-standard embedding models~\cite{Braun:2005ux}--\cite{Anderson:2009mh}, where $V$ is a more general bundle over $X$. Only a relatively small number of models exhibiting a realistic massless spectrum have been constructed in this way~\cite{Bouchard:2005ag,Braun:2005ux,Anderson:2009mh,Braun:2011ni}, reflecting the considerable technical problems associated with vector bundles on smooth Calabi-Yau manifolds. They are complemented by models found in related heterotic constructions such as those based on orbifolds~\cite{Buchmuller:2005jr,Buchmuller:2006ik,Lebedev:2006kn,Kim:2007mt,Lebedev:2007hv,Lebedev:2008un,Nibbelink:2009sp,Blaszczyk:2009in,Blaszczyk:2010db,Kappl:2010yu}, on the free fermionic strings~\cite{Assel:2009xa,Christodoulides:2011zs,Cleaver:2011ir}, and on Gepner models \cite{Maio:2011qn,GatoRivera:2009yt,GatoRivera:2010xn}. Overall, it is fair to say that the number of quasi-realistic heterotic models, as counted by the number of underlying GUT models, has been relatively small. 

This situation has changed with the results published in Ref.~\cite{Anderson:2011ns} where 200 heterotic Calabi-Yau GUT models were presented. By verifying a number of general criteria it was shown that each of these models leads to heterotic standard models upon suitable quotienting by discrete symmetries and including Wilson lines. This progress has been possible for two main reasons. Firstly, rather than following a ``model building approach" by trying to fine-tune individual models for the right phenomenological properties, systematic scans, using methods of computational algebraic geometry, have been performed over large classes of models and unsuitable candidates have been successively filtered out. The considerable mathematical and computational tools necessary for such systematic scans have been built up over a number of years~\cite{Anderson:2007nc,Anderson:2008uw,Gray:2008zs,cicypackage,Anderson:2009ge,Anderson:2009mh,Gray:2009fy,Blumenhagen:2010pv,Rahn:2011jw}. The second reason is related to the nature of the vector bundles $V$ used in the construction. Previous model building attempts~\cite{Distler:1987ee,Kachru:1995em,Bouchard:2005ag,Braun:2005ux,Braun:2005bw,Braun:2005zv,Braun:2005nv,Bouchard:2006dn,Blumenhagen:2006ux,Blumenhagen:2006wj,Anderson:2007nc,Anderson:2008uw,Anderson:2009mh} have mostly focused on vector bundles $V$ with a non-Abelian structure group. However, once we move away from the standard embedding, the complexity of the constructions rather motivates studying the simplest bundle choices, that is, bundles with Abelian structure groups. Using such Abelian bundles is one of the key ideas underlying the work in Ref.~\cite{Anderson:2011ns} and the present paper. The technical simplifications which arise in the case of Abelian bundles greatly facilitate the systematic scanning and the construction of a sizeable number of promising models. 

The $200$ models given in Ref.~\cite{Anderson:2011ns} were essentially constructed at the ``upstairs" GUT level. The structure group of the bundle $V$ on the Calabi-Yau manifolds $X$ was chosen to be $S(U(1)^5)\subset E_8$ so that the low-energy gauge group is $SU(5)\times S(U(1)^5)$, with the additional $U(1)$ symmetries being Green-Schwarz anomalous in most cases. The GUT matter spectrum for all models consists of $3|\Gamma|$ ${\bf 10}$ and $\bar{\bf 5}$ multiplets, some number of ${\bf 5}$--$\bar{\bf 5}$ vector-like pairs and a number of $SU(5)$ singlets. Here $|\Gamma|$ is the order of a freely-acting discrete symmetry $\Gamma$ on $X$. It is clear that quotienting these models by $\Gamma$ and including Wilson lines in order to break $SU(5)$ can lead to a low-energy theory with the standard model group (times anomalous $U(1)$ symmetries with massive associated gauge bosons) and three families of quarks and leptons. Further, provided certain constraints on the number of ${\bf 5}$--$\bar{\bf 5}$ pairs hold one can ensure that all Higgs triplets can be projected out and at least one pair of Higgs doublets can be kept. In Ref.~\cite{Anderson:2011ns} it was shown that these constraints are indeed satisfied, so that all $200$ models lead to heterotic standard models without any exotic fields charged under the standard model group. 

In the present paper we go one step further and construct the downstairs standard models which result from the 200 GUT models of Ref.~\cite{Anderson:2011ns} explicitly. We compute the complete spectrum, including Higgs multiplets and gauge singlet fields, for each model, thereby determining the $S(U(1)^5)$ charges for all multiplets. Taking into account all different choices of quotienting the bundle and including the Wilson lines, this leads to tens of thousands of downstairs models. In this paper we focus on the four-dimensional spectrum of particles and operators and, hence, we identify models which descend from the same upstairs theory if they lead to the same four-dimensional fields.  After removing these and some other redundancies we find about $400$ models, each with the standard model gauge group times $S(U(1)^5)$, precisely three families of quarks and leptons, between one and three pairs of Higgs doublets and no exotic fields charged under the standard model group of any kind. In addition, we have a number of standard model singlet fields, $S^\alpha$, which are charged under $S(U(1)^5)$. To the best of our knowledge, this is the largest set of string models with precisely the standard model spectrum found to date. Details of all models can be found in the standard model database~\cite{database}.

From a 10-dimensional point of view the singlet fields $S^\alpha$ can be interpreted as bundle moduli, where vanishing vacuum expectation values for $S^\alpha$ correspond to the original Abelian gauge bundle and non-zero vacuum expectation values indicate a deformation to a bundle with non-Abelian structure group. We would like to stress that, despite the presence of the additional $S(U(1)^5)$ symmetry, there is no problem with additional massless vector bosons. For most models, all additional $U(1)$ symmetries are Green-Schwarz anomalous~\cite{gsw,Dine:1987xk,Lukas:1999nh,Blumenhagen:2005ga,Anderson:2009nt,Anderson:2009sw,Anderson:2010ty} and, hence, the associated gauge bosons are super-heavy. If one of the $U(1)$ symmetries remains non-anomalous (and the associated gauge boson is massless), as happens in some cases, it can easily be spontaneously broken by turning on vacuum expectation values for the singlet fields $S^\alpha$. As discussed above, this corresponds to deforming the gauge bundle to a one with a non-Abelian structure group. 

Despite the similarity of their low-energy field content our models are distinct in a number of ways. Most importantly, the $S(U(1)^5)$ charges of matter and Higgs multiplets can vary between models. In addition, the numbers and $S(U(1)^5)$ charges of the singlets $S^\alpha$ are model-dependent, as is the number of Higgs doublet pairs. Taking this into account, we find $76$ different spectra among the $400$ models. However, even models with an identical four-dimensional spectrum have a different higher-dimensional origin and can, therefore, be expected to differ at a more sophisticated level, for example in the values of their coupling constants. For this reason, we have kept all $400$ models in our database~\cite{database}. 

Our models fall within a general class of four-dimensional $N=1$ supergravity theories obtained from heterotic line bundle compactifications which we would like to refer to as {\em line bundle standard models}. From a four-dimensional point of view, these models are characterized by an NMSSM-type spectrum (however, with generally many rather than just one singlet field), the presence of an additional Green-Schwarz anomalous $S(U(1)^5)$ symmetry and a specific pattern of charges under this symmetry. The presence of these additional symmetries constrains the allowed operators in the four-dimensional theory and thereby facilitates the study of phenomenological properties beyond the computation of the matter spectrum. They can be phenomenologically helpful, for example by forbidding proton decay violating operator, or phenomenologically dangerous, for example if they force all Yukawa couplings to vanish. A wide range of phenomenological issues, including flavour physics, proton decay, the $\mu$ term, R-parity violation and neutrino masses can be addressed in this way. In Ref.~\cite{Anderson:2011ns} this was carried out for a particular example. In the present paper, we compute the allowed set of operators in the four-dimensional theory for all $400$ models and the results are listed in the database~\cite{database}. These results allow for a more detailed study of the models' phenomenology and we discuss a number of generic features based on these results. For example, we find that $45$ of our models allow for an up Yukawa matrix with non-vanishing rank, before switching on singlet vacuum expectation values. For about $200$ of our models, all dimension four and five proton-decay violating operators are forbidden by the $S(U(1)^5)$ symmetry. We have $262$,  $77$ and $63$ models, respectively, with one, two and three pairs of Higgs doublets. Requiring precisely one pair of Higgs doublets, the absence of all dimension four and five proton decay violating operators, an up-Yukawa matrix with non-zero rank and no massless $U(1)$ vector boson (in the absence of singlet VEVs), $13$ models remain. 

Because of the somewhat technical nature of the underlying 10-dimensional construction we have split the paper into two parts which can largely be read separately. The first part, which consists of sections 2 and 3, describes heterotic line bundle models purely from the perspective of the four-dimensional $N=1$ supergravity theory. In section 2, we set up the general structure of these four-dimensional models. Section 3 presents an example model from the database~\cite{database}, in order to discuss various phenomenological issues and explain the structure of the data files. We end the section with an overview of basic phenomenological properties among our $400$ standard models. 
The remainder of the paper describes the construction of the models starting with the 10-dimensional theory. In section 4, we set up the general formalism for heterotic Calabi-Yau compactifications in the presence of vector bundles with split structure groups. We also explain our scanning criteria and procedure in general. Section 5 describes our specific arena for the construction of models, that is, complete intersection Calabi-Yau manifolds (CICYs) and line bundles thereon, as well as details of the scanning procedure. A number of specific issues which arise in heterotic Calabi-Yau models with split bundles is discussed in Section 6. Our summary and outlook is presented in Section 7. Appendices A and B contain additional technical information on the construction of equivariant structures and the computation of equivariant cohomology.

\section{Line bundle standard models}
\label{sec:lbsms}
In this section, we introduce the general class of four-dimensional $N=1$ supergravity theories with a standard model spectrum, derived from heterotic line bundle compactifications on Calabi-Yau manifolds. We will refer to this class of supergravities as line bundle standard models. This sets the scene for the discussion in Section 3, where we present an explicit example from our standard model data base and a general phenomenological overview of our models. In addition, this class of supergravities provides a general framework for string phenomenology within a purely four-dimensional setting. Indeed, we expect many more line bundle models to exist than are currently available in our database~\cite{database}, constructed by considering more general line bundles and other Calabi-Yau manifolds. All of these models will be described by a supergravity of the type introduced below. 

\subsection{The gauge group}
The gauge group of line bundle standard models is given by the standard model group $G_{\rm SM}=SU_c(3)\times SU_W(2)\times U_Y$ times the additional gauge symmetry $\anU=S(U(1)^5)$. We can think of the elements of $\anU$ as given by five phases $(e^{i\eta^1},\ldots ,e^{i\eta^5})$ subject to the ``determinant one" condition $\sum_{a=1}^5\eta^a=0$. Although $\anU\cong U(1)^4$ it will be more convenient for our purposes to work with $\anU$ rather than $U(1)^4$. Irreducible $\anU$ representations can be labelled by an integer vector ${\bf q}=(q_a)_{a=1,\ldots ,5}$. However, due to the determinant one condition two such vectors, ${\bf q}$ and $\tilde{\bf q}$ refer to the same representation and, hence, have to be identified iff
\begin{equation}
 {\bf q}-\tilde{\bf q}\in \mathbb{Z}(1,1,1,1,1)\; . \label{qid}
\end{equation} 
In particular, this means that a four-dimensional operator is $\anU$ invariant precisely if the five entries in its charge vector are identical. All standard model multiplets carry $\anU$ charges which follow a specific pattern originating from the underlying string construction. This structure of charges will be introduced below. 

We stress that the four gauge bosons associated to $\anU$ do not cause a phenomenological problem. In most cases, all $U(1)$ symmetries are Green-Schwarz anomalous and, hence, the gauge bosons receive a super-heavy Stueckelberg mass. In cases where some of the $U(1)$ symmetries are non-anomalous masses for the associated gauge bosons can be generated by spontaneously symmetry breaking through VEVs of standard model singlet fields. This will be discussed in more detail in the section on $U(1)$ vector boson masses below. 

\subsection{The matter field sector}
Matter fields transform linearly under $\anU$, that is,
\begin{equation}
 \Phi\rightarrow\exp(i{\bf q}.{\boldsymbol\eta})\Phi\label{Phitrafo}
\end{equation} 
for a matter field $\Phi$ with $\anU$ charge ${\bf Q}(\Phi)={\bf q}$. Although there is no four-dimensional $SU(5)$ GUT symmetry it turns out that the $\anU$ charge is always identical for all fields in a given $SU(5)$ multiplet. For this reason, it is useful to combine the three standard model families into $SU(5)$ multiplets and introduce the notation $({\bf 10}^p)=(Q^p,u^p,e^p)$ and $(\bar{\bf 5}^p)=(d^p,L^p)$, where $p,q,\ldots =1,2,3$ are family indices. Their pattern of $\anU$ charges is given by
\begin{equation}
 {\bf Q}({\bf 10}^p)={\bf e}_{a_p}\; ,\quad {\bf Q}(\bar{\bf 5}^p)={\bf e}_{b_p}+{\bf e}_{c_p}\; , \label{famcharges}
\end{equation} 
where $a_p,b_p,c_p=1,\ldots ,5$ and $b_p<c_p$. Here ${\bf e}_a$ denotes the $a^{\rm th}$ standard unit vector in five dimensions. Hence, ${\bf 10}$ families have charge one under precisely one of the five $U(1)$ symmetries in $\anU$, while $\bar{\bf 5}$ multiplets have charge one with respect to two of the $U(1)$ symmetries.  Apart from these rules, the precise pattern of charges across the three families is model dependent. For example, for the three ${\bf 10}$ families, there are models with all three $\anU$ charges the same, two charges the same and the third one different or all three charges different. To specify explicit models it will be convenient to introduce a simple notation for the $\anU$ charge. We do this by adding a charge label as a subscript to the multiplet's name so that, for example ${\bf 10}_2$ denotes a ${\bf 10}$ multiplet with charge ${\bf e}_2=(0,1,0,0,0)$ and $\bar{\bf 5}_{1,4}$ denotes a $\bar{\bf 5}$ multiplet with charge ${\bf e}_1+{\bf e}_4=(1,0,0,1,0)$. 

In addition, we have one (or, in some cases, more than one) pair of Higgs doublets $H$, $\bar{H}$ with $\anU$ charges of the type
\begin{equation}
 {\bf Q}(H)={\bf e}_{h}+{\bf e}_{g}\; ,\quad {\bf Q}(\bar{H})=-{\bf e}_{\bar{h}}-{\bf e}_{\bar{g}}\; ,
\end{equation}
where $h,g,\bar{h},\bar{g}=1,\ldots ,5$ and $h<g$, $\bar{h}<\bar{g}$. As before, we attach the $\anU$ charge as a subscript so that, for example, a down Higgs $H_{2,3}$ has charge ${\bf e}_2+{\bf e}_3=(0,1,1,0,0)$ and an up-Higgs $\bar{H}_{3,5}$ has charge $-{\bf e}_3-{\bf e}_5=(0,0,-1,0,-1)$. 

Finally, line bundle standard models come with standard model singlet fields, which we denote by $S^\alpha$. Their number is model-dependent and, for typical examples, varies between a few and a few $\times 10$. Their $\anU$ charges have the form
\begin{equation}
 {\bf Q}(S^\alpha)={\bf e}_{d_\alpha}-{\bf e}_{f_\alpha}\; ,
\end{equation}
where $d_\alpha ,f_\alpha =1,\ldots ,5$. Following the convention for the other fields we append this charge as a subscript so that, for example, $S_{2,5}$ has charge ${\bf e}_2-{\bf e}_5=(0,1,0,0,-1)$ and $S_{4,1}$ has charge ${\bf e}_4-{\bf e}_1=(-1,0,0,1,0)$. As mentioned in the introduction, from a 10-dimensional point of view, these singlet fields can be interpreted as gauge bundle moduli. Vanishing VEVs for all singlets correspond to Abelian gauge bundles while non-vanishing VEVs indicate a deformation to non-Abelian structure groups. The singlets also play an important role from the viewpoint of the four-dimensional theory since they always carry a non-trivial $\anU$ charge. This means that non-vanishing singlet VEVs can spontaneously break $U(1)$ symmetries in $\anU$, thereby giving mass to the vector bosons associated to non-anomalous $U(1)$ factors which have not received a mass from the Stueckelberg mechanism. 

In summary, the matter spectrum of line bundle standard models is that of a generalized NMSSM, typically with a number of singlet fields rather than just a single one, and with a specific pattern of ${\cal J}=S(U(1)^5)$ charges, as explained above. 
\subsection{The moduli sector}
The gravitational moduli of the models consist of the dilaton, ${\cal S}=s+i\sigma$, a certain number of Kahler moduli, denoted by $T^i=t^i+2i\chi^i$, and complex structure moduli generically denoted by $Z$. All the moduli are singlets under the standard model group. The complex structure moduli are also singlets under the $U(1)$ symmetries in $\anU$ but the Kahler moduli and the dilaton have non-linear transformations, acting an their respective axionic components $\chi^i$ and $\sigma$ as
\begin{equation}
 \delta\chi^i = -k^i_a\eta^a\; ,\quad \delta\sigma =-2k^i_a\beta_i\eta^a\; . \label{chitrafo}
\end{equation} 
Here, ${\bf k}_a=(k^i_a)$ and ${\boldsymbol\beta}=(\beta_i)$ are numbers which are fixed for a given string construction and can be determined from the underlying topology, as will be discussed in Section \ref{mrsection4sir}. The special unitarity of the gauge group ${\cal J}=S(U(1)^5)$ means that the vectors ${\bf k_a}$ are subject to the constraint
\begin{equation}
 \sum_{a=1}^5{\bf k}_a=0\; . \label{sumk}
\end{equation} 
The 10- or 11-dimensional origin of our theories implies certain constraints on the moduli fields which are necessary for the validity of the four-dimensional effective theory. In particular, it is necessary that
\begin{equation}
 t^i\gg1\; ,\quad \frac{\beta_it^i}{s}\ll 1\; . \label{modcons}
\end{equation}
The first of these constraints ensures that the internal Calabi-Yau volume and the volume of cycles therein is sufficiently large for the supergravity approximation to be valid. The second constraint is necessary for the strong coupling expansion~\cite{Witten:1996mz,Lukas:1998hk} of the 11-dimensional theory to be valid.

In addition, the model can have moduli associated to the hidden $E_8$ sector and to five-branes (if present in the construction), all of which are standard model singlets. They will not play an essential role for the subsequent discussion.

\subsection{The effective action} \label{tea}
We begin by writing down the generic form for the superpotential which we split up as
\begin{equation}
 W=W_{\rm Y}+W_{\rm R}+W_5+W_{\rm sing}+W_{\rm np}\; . \label{W}
\end{equation}
The first four terms are perturbative while $W_{\rm np}$ contains the non-perturbative contributions. The standard Yukawa couplings and the $\mu$-term are contained in $W_{\rm Y}$,  $W_{\rm R}$ consists of the R-parity violating terms and $W_5$ consists of the order five terms in standard model fields. The pure singlet field terms are collected in $W_{\rm sing}$. Schematically, these perturbative parts can be written as
\begin{eqnarray}
 W_{\rm Y}&=&\mu H\bar{H}+Y^{(d)}_{pq}H\bar{\bf 5}^p{\bf 10}^q+Y^{(u)}_{pq}\bar{H}{\bf 10}^p{\bf 10}^q\label{WYuk}\\
W_{\rm R}&=&\rho_p \bar{H}L^p+\lambda_{pqr}\bar{\bf 5}^q\bar{\bf 5}^q{\bf 10}^r\label{WR}\\
W_5&=&\lambda_{pqrs}'\bar{\bf 5}^p{\bf 10}^q{\bf 10}^r{\bf 10}^s\label{Wprot5}\\
W_{\rm sing}&=&\tau_{\alpha\beta\gamma}S^\alpha S^\beta S^\gamma\; .
 \end{eqnarray} 
For simplicity, we have expressed the operators in terms of GUT multiplets, wherever possible. Since the $U(1)$ charges in $\anU$ commute with $SU(5)$ this will be sufficient to discuss the pattern implied by $\anU$-invariance, which is our main purpose. It should, however, be kept in mind that the precise values of the allowed couplings will, in general, break $SU(5)$. This means, for example, that the standard $SU(5)$ GUT relation between tau and bottom Yukawa couplings may not be satisfied. All couplings above should be thought of as functions of moduli. As usual, they cannot depend on the dilaton, ${\cal S}$, and the Kahler moduli $T^i$ thanks to their axionic shift symmetries (some of which are even gauged according to Eq.~\eqref{chitrafo}). However, they are, in general, functions of the complex structure moduli $Z$ and the singlet fields (bundle moduli) $S^\alpha$. 

In this paper, for the most part, we will be interested in studying the theory for the locus in moduli space where all singlet fields are small, so $|S^\alpha|\ll 1$. From a 10-dimensional point of view this means we are considering gauge bundles with Abelian structure group or small non-Abelian deformations thereof. On this locus, all couplings above can be expanded in powers of $S^\alpha$ around the ``Abelian locus" $S^\alpha=0$. For example, for the $\mu$-term we can write~\footnote{For models with multiple pairs of Higgs doublets the $\mu$-term of course generalizes to a matrix of $\mu$-terms.}
\begin{equation}
 \mu = \mu_0+\mu_{1,\alpha}S^\alpha+\mu_{2,\alpha\beta}S^\alpha S^\beta+\dots\; , \label{muexp}
\end{equation} 
and similarly for all other couplings. In general, the expansion coefficients $\mu_0$, $\mu_{1,\alpha}$, etc.~should still be considered functions of the complex structure moduli. Their pattern is restricted by the $\anU$ charges of the standard model fields and the singlet fields $S^\alpha$ and it is this structure which we will mainly analyze in the following. Also note that the zeroth order $\mu$-term, $\mu_0$, in Eq.~\eqref{muexp} vanishes even  if the Higgs pair is vector-like under $\anU$ since all our models have an exactly massless Higgs pair at the Abelian locus $S^\alpha=0$.

In the rest of the paper, we will not consider the non-perturbative superpotential $W_{\rm np}$ but a few remarks concerning its structure may be in order. Generally, one expects two types of non-perturbative effects to contribute: string instanton effects leading to terms of the form $P(Z,S^\alpha)\exp (-n_iT^i)$, and gaugino condensation leading to terms of the form $Q(Z,S^\alpha)\exp(-c({\cal S}\pm\beta_iT^i)$. Here, $c$ and $n_i$ are positive constants (related to the beta function of the condensing gauge group and the instanton number, respectively) and $P$, $Q$ are functions, typically rational, of the moduli $Z$ and $S^\alpha$. The main point is that the presence of the $U(1)$ gauge symmetries in $\anU$ significantly constrains the allowed non-perturbative terms, in view of the transformations~\eqref{chitrafo} and \eqref{Phitrafo}. Specifically, the phase change of the non-perturbative exponentials due to the axion transformations~\eqref{chitrafo} has to be cancelled by the phase change of the pre-factors $P$, $Q$ due to the linear transformations of the singlet fields $S^\alpha$. In Ref.~\cite{Anderson:2011cza} this has been analyzed for the special case when singlet fields $S^\alpha$ are absent. The more general case with singlets remains to be considered in detail and this will clearly be central for the discussion of moduli stabilization and supersymmetry breaking in heterotic line bundle models. 

Let us now move on to the general structure of the Kahler potential. As usual, it can be written as a sum 
\begin{equation}
 K=K_{\rm mod}+K_{\rm mat}
\end{equation}
of the moduli superpotential $K_{\rm mod}$ and the matter superpotential $K_{\rm mat}$. For the former, we have
\begin{equation}
 K_{\rm mod}=-\ln({\cal S}+\bar{{\cal S}})-\ln (\kappa)+K_{\rm cs}+\dots\; , \label{Kmod}
\end{equation}
where $K_{\rm cs}$ is the standard special geometry Kahler potential for complex structure moduli~\cite{Candelas:1990pi},  and the dots stand for contributions from other moduli. The quantity $\kappa$ is defined as
\begin{equation}
 \kappa=d_{ijk}t^it^jt^k\; , \label{kappadef}
\end{equation}
with numbers $d_{ijk}$. From a 10-dimension viewpoint $\kappa$ is proportional to the Calabi-Yau volume and $d_{ijk}$ are the triple intersection numbers of the Calabi-Yau manifold. It is also useful to introduce the Kahler metric for the $T^i$ moduli which follows from the above Kahler potential. It is given by
\begin{equation}
 G_{ij}\equiv-\frac{1}{4}\frac{\partial^2}{\partial t^i\partial t^j}\ln\kappa = -\frac{3}{2}\left(\frac{\kappa_{ij}}{\kappa}-\frac{3}{2}\frac{\kappa_i\kappa_j}{\kappa^2}\right)\; .
\end{equation}
with $\kappa_i=d_{ijk}t^jt^k$ and $\kappa_{ij}=d_{ijk}t^k$.

The matter field Kahler potential has the structure
\begin{eqnarray}
 K_{\rm mat}&=&K^{({\bf 10})}_{p\bar{q}}{\bf 10}^p{\bf 10}^{\bar{q}\dagger}+K^{(\bar{\bf 5})}_{p\bar{q}}\bar{\bf 5}^p\bar{\bf 5}^{\bar{q}\dagger}+
                      K^{(u)}\bar{H}\bar{H}^{\dagger}+K^{(d)}HH^{\dagger}\nonumber\\
                      &&+\left(\tilde{\mu}H\bar{H}+\tilde{\rho}_pL^p\bar{H}+\hat{\rho}_pL^pH^{\dagger}+{\rm c.c}\right)+K_{\rm sing}\; ,\label{Kmatter}
\end{eqnarray} 
where $K_{\rm sing}$ is the singlet superpotential which depends on the singlets $S^\alpha$ and their conjugates but not on the other matter fields. The couplings in $K_{\rm mat}$ should be considered as functions of the moduli, more specifically of ${\cal S}+\bar{\cal S}$, $T^i+\bar{T}^i$, $Z$, $Z^\dagger$, $S^\alpha$ and $S^{\alpha\dagger}$. As before, for small $S^\alpha$ we can expand all couplings around the locus $S^\alpha=0$, for example
\begin{equation}
  K^{(u)}=K^{(u)}_0+\left(K^{(u)}_{1,\alpha}S^\alpha +{\rm c.c.}\right)+\cdots\; ,
\end{equation}  
and similarly for the other couplings. The expansion coefficients are still functions of the other moduli and, as for the superpotential, they are restricted by $\anU$ invariance. 

Some general remarks about the constraints implied by $\anU$ invariance are in order. Of course we know that $\anU$ non-invariant terms must be absent from the action. A $\anU$ invariant term will typically be present with a coupling which is of order one for generic values of the complex structure moduli. However, it is still possible that this coupling vanishes for specific values of the complex structure moduli. The term might even be forbidden altogether for reasons unrelated to the $\anU$ symmetry, for example, because of the presence of an additional discrete symmetry in the model. We can, therefore, safely draw conclusions from the absence of certain terms due to $\anU$ non-invariance, but we have to keep this limitation in mind when we rely on the presence of $\anU$-invariant operators. In principle, we can improve on this point since many of the couplings can be explicitly computed from the underlying string theory~\cite{Anderson:2009ge}. This task is beyond the scope of the present paper and will be addresses in future publications. 

The gauge kinetic function for the standard model group is universal, as is usually the case in heterotic theories, and given by
\begin{equation}
 f={\cal S}+\beta_iT^i\; , \label{f}
\end{equation}
with the topological numbers $\beta_i$ identical to the ones which appear in the transformations~\eqref{chitrafo} of the axions. In view of these, the gauge kinetic function transforms non-trivially under a the $U(1)$ symmetries in $\anU$, namely
\begin{equation}
 \delta f=-4ik_a^i\beta_i\eta^a\; . \label{df}
\end{equation} 
As we will see, this non-trivial classical variation cancels the mixed ${\cal J}G_{\rm SM}G_{\rm SM}$ triangle anomaly in a four-dimensional realization of the Green-Schwarz mechanics. The gauge kinetic function for the $U(1)$ vector fields in $\anU$ is given by~\cite{Lukas:1999nh}
\begin{equation}
 f_{ab}=f\delta_{ab}+\frac{2}{3}d_{ijk}k_a^jk_b^kT^i\; .  \label{fab}
\end{equation}
Note that the second term represents a kinetic mixing between the $U(1)$ symmetries. In the presence of anomalous $U(1)$ symmetries in the hidden sector this kinetic mixing becomes more complicated and involves cross terms between hidden and observable $U(1)$ symmetries. Since we are here focusing on the observable matter field sector we will not consider this explicitly. The general form of the gauge kinetic function, including hidden-observable mixing, can be found in Ref.~\cite{Lukas:1999nh}. The variation of~\eqref{fab} leads to a cancellation of the ${\cal J}^3$ triangle anomaly.  

This concludes our general set-up of heterotic line bundle models. It remains to discuss a number of generic features of these theories which are all related to the presence of the additional $U(1)$ symmetries in $\anU$.

\subsection{D-terms}
\label{sec:Dterms}
In this subsection we would like to discuss the D-terms associated to the $U(1)$ gauge symmetries in $\anU$. They can be computed from the linear matter fields transformations~\eqref{Phitrafo} and the non-linear transformations~\eqref{chitrafo} of the dilaton and the T-moduli using standard supergravity methods~\cite{Wess:1992cp}. Explicitly they are given by
\begin{equation}
D_a=\frac{3 k_a^i\kappa_i}{\kappa}+\frac{\beta_ik^i_a}{s}-\sum_{P,\bar{Q}}q_{Pa}K_{P\bar{Q}}C^P\bar{C}^{\bar Q}\; . \label{Dterm}
\end{equation} 
Here, $C^P$ collectively denote all matter fields with $\anU$ charges $q_{Pa}$ and $K_{P\bar{Q}}$ is their Kahler metric as computed from Eq.~\eqref{Kmatter}. In particular, these matter fields include the singlets $S^\alpha$. Since the gauge group $\anU$ is special unitary there are in fact only four-independent D-terms. Indeed, as a consequence of Eq.~\eqref{sumk} and the structure of the matter field charges the above D-terms satisfy the relation
\begin{equation}
 \sum_{a=1}^5D_a=0\; . \label{sumD}
\end{equation}
For a supersymmetric vacuum at or near the locus $S^\alpha=0$ we need to solve the D-term equations $D_a=0$ along with the F-term equations which follow from the singlet superpotential $W_{\rm sing}$ in \eqref{W}. In general, this requires specific knowledge of the singlet superpotential and the matter field part in the D-term~\eqref{Dterm}. For a given model in our database both will normally be highly constrained by $\anU$ invariance so that this analysis can be carried out explicitly on a case-by-case basis. However, since we think of our models as being defined near $S^\alpha=0$ we should first ensure that a supersymmetric vacuum exists at this Abelian locus. In this case, the F-term equations for $S^\alpha$ are automatically satisfied and the matter field contributions to the D-terms vanish. In other words, we have to ensure that the FI terms, corresponding to the first two terms in Eq.~\eqref{Dterm}, vanish. Evidently, this imposes restrictions on the dilaton and the T-moduli. To this end, let us introduce the corrected T-moduli $\tilde{t}_i=3 \kappa_i/\kappa+\beta_i/s$. Note that in view of the constraints~\eqref{modcons} on the moduli, the second term in this definition is indeed a small correction. Then the D-term equations can be written as
\begin{equation}
  D_a=k_a^i\tilde{t}_i=0\; .
\end{equation}  
A non-trivial solution to these equations exists only if
\begin{equation}
 (\mbox{number of lin.~independent }{\bf k}_a)<(\mbox{number of T-moduli})\; . \label{numcons}
\end{equation}
Hence, for models with less than five Kahler moduli further linear dependencies, in addition to \eqref{sumk}, must exist between the charge vectors ${\bf k}_a$. This implies a significant model-building constraint for models with a small number of Kahler moduli. \\

\subsection{Green-Schwarz anomaly cancellation}
\label{sec:gs4d}
The $U(1)$ symmetries in $\anU$ are generically anomalous in our models. In particular, this means that the mixed ${\cal J}G_{\rm SM}G_{\rm SM}$ triangle anomalies between a $\anU$ gauge boson and two standard model gauge bosons as well as the cubic ${\cal J}^3$ anomaly between three $\anU$ gauge bosons are typically non-vanishing. The Green-Schwarz mechanism, in its four-dimensional version, implies that these triangle anomalies are cancelled due to the non-trivial $\anU$-transformations~\eqref{f}, \eqref{fab} of the gauge-kinetic functions. 

We begin, by discussing this explicitly for the mixed ${\cal J}G_{\rm SM}G_{\rm SM}$ anomalies. Using the charges~\eqref{famcharges} for the ${\bf 10}$ and $\bar{\bf 5}$ families the coefficients of these triangle anomalies are given by
\begin{equation} 
{\bf A}=\sum_{p=1}^3\left(3{\bf e}_{a_p}+{\bf e}_{b_p}+{\bf e}_{c_p}\right)\; . \label{trianom}
\end{equation}
For these to be cancelled by the transformation~\eqref{df} of the gauge kinetic function we have to require that
\begin{equation}
  \left(A_a-k^i_a\beta_i\right)_{a=1,\ldots ,5}\in \mathbb{Z}(1,1,1,1,1)\; . \label{anomcons}
\end{equation}
For the models in our database these relations are automatically satisfied due to the Green-Schwarz mechanism in the underlying 10-dimensional theory. However, from a bottom-up point of view this constitutes a significant constraint, relating the charge choices for the matter fields and the moduli fields with the parameters $\beta_i$ which determine the size of the threshold correction to the gauge kinetic function. 

Similarly, the ${\cal J}^3$ triangle anomaly must be cancelled by the variation of the $U(1)$ gauge kinetic functions~\eqref{fab}. This leads to constraints analogous to Eq.~\eqref{anomcons} which, however, also depend on the spectrum of singlet fields $S^\alpha$. For this reason they are of less practical importance and we will not present them explicitly. 

\subsection{Masses of $U(1)$ gauge bosons}
\label{sec:U1masses}
The mass terms for the $U(1)$ vector bosons arise from the kinetic terms for the axions $\sigma$ and $\chi^i$ as a consequence of the non-linear transformations~\eqref{chitrafo} and, for non-vanishing VEVs for the singlets $S^\alpha$, also from the kinetic terms of those fields. At the Abelian locus, $S^\alpha=0$, only the former contribution is present and results in a mass matrix
\begin{equation}
 M_{ab}={\bf k}_a^T\tilde{G}{\bf k}_b\;\mbox{ where }\;\tilde{G}_{ij}=G_{ij}+\frac{\beta_i\beta_j}{4s^2}\label{U1masses}
\end{equation}
is the corrected Kahler metric for the T-moduli. Since $\tilde{G}_{ij}$ is non-degenerate this means that the number of massless $U(1)$ vector bosons at the locus $S^\alpha=0$ is given by
\begin{equation}
 (\mbox{number of massless }U(1)\mbox{ vector bosons})=4-\mbox{rank} (k_a^i)\; . \label{numU1}
\end{equation}
Such a massless linear combination of vector bosons, characterized by a vector $v^a$ satisfying ${\bf k}_av^a=0$, corresponds to a non-anomalous $U(1)$ symmetry, as can be seen, in the case of the mixed anomaly, from Eq.~\eqref{anomcons}. Combining the above result with Eq.~\eqref{numcons} we learn that
\begin{equation}
 (\mbox{number of massless }U(1)\mbox{ vector bosons})>4-(\mbox{number of T-moduli})\; .
\end{equation} 
In particular, for models with less than five Kahler moduli, there necessarily exists at least one massless $U(1)$ vector boson at the Abelian locus. On the other hand, for five or more Kahler moduli all $U(1)$ vector bosons will be generically massive. 

Non-anomalous $U(1)$ symmetries can of course be easily broken spontaneously, thereby giving masses to the associated vector bosons, by switching on $S^\alpha$ VEVs. For this reason, there is no serious phenomenological problem with the presence of additional massless $U(1)$ symmetries at the Abelian locus and we have included such models in our database. In a detailed analysis it has of course to be checked that this spontaneous breaking is consistent with supersymmetry, that is, that it can be achieved for vanishing F- and D-terms. 

\section{The model database}
\label{sec:database}
After this general set-up we will now present the line bundle standard models from heterotic compactifications which are accessible from the database~\cite{database}. This will be done mainly from the viewpoint of the four-dimensional effective theories, following the set-up of the previous section. The underlying 10-dimensional construction will be explained in the following section. We begin by presenting one specific example model from the database. There is no implication that this particular model is phenomenologically favoured or even viable. It has merely been chosen as a useful example to explain the contents of the database and to illustrate the possible phenomenological applications of heterotic line bundle models. 

In the second part of this section, we will discuss the distribution of basic phenomenological properties in our database. For example, we will count the number of models with one, two and three pairs of Higgs doublets, the number of models with vanishing dimension four and five proton-decay inducing terms and similar properties. 

\subsection{An example model}
\label{sec:example}
We will now present an example model from the database~\cite{database}, namely model number 7 on the Calabi-Yau manifold with number 6732. 
First we discuss the gravitational sector and then move on to the matter fields and the detailed spectrum of allowed operators in the four-dimensional effective theory.
\subsubsection{The gravitational sector}
\begin{figure}[!h]
\centering
\includegraphics[width=14cm]{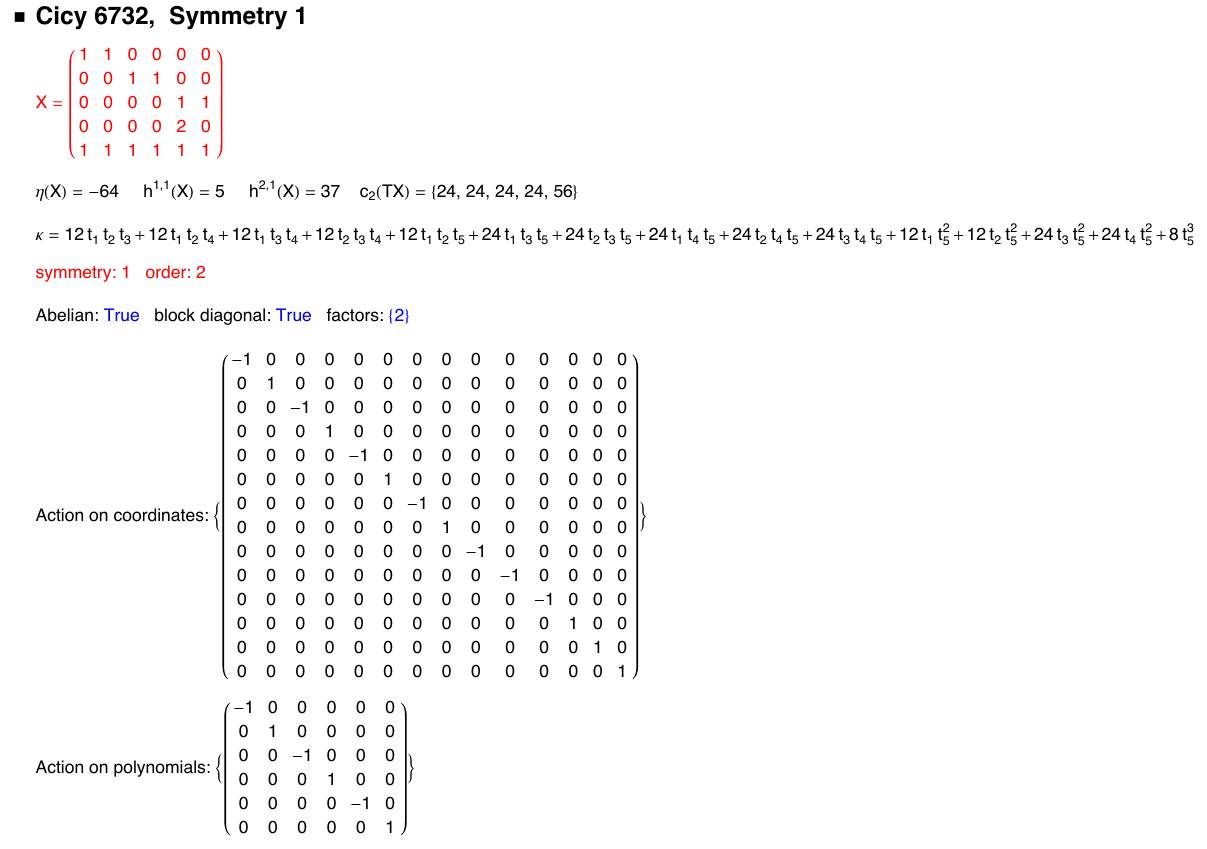}
\parbox{6in}{\caption{\it\small Example database entry for complete intersection Calabi-Yau three-fold 6732.}\label{fig:cicy6732}}
\end{figure}

\begin{figure}[!h]
\centering
\includegraphics[width=14cm]{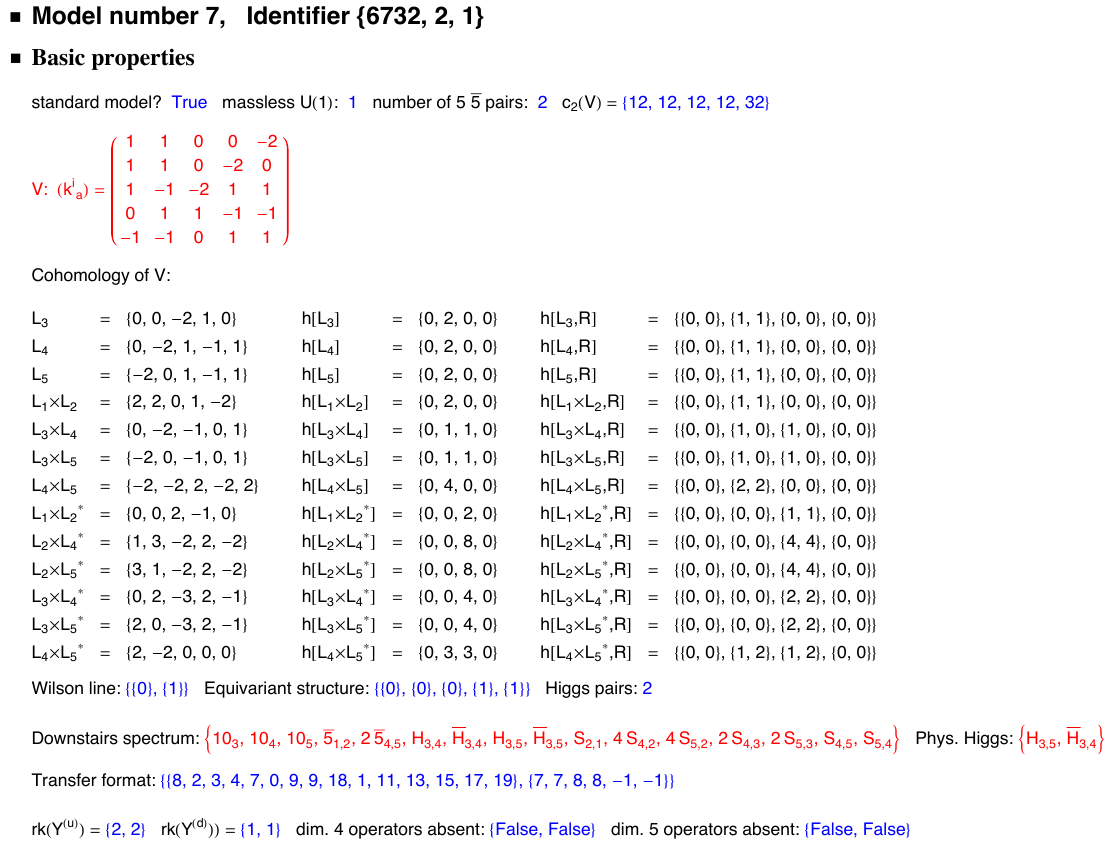}
\parbox{6in}{\caption{\it\small Example database entry specifying basic properties of a line bundle standard model on the three-fold in Fig.~\ref{fig:cicy6732}.}\label{fig:model7_1}}
\end{figure}

The database entry for the Calabi-Yau manifold underlying our example model is shown in Fig.~\ref{fig:cicy6732}.
The data given in the figure defines a Calabi-Yau three fold $X$ with a freely-acting symmetry group $\Gamma$. The actual Calabi-Yau manifold underlying the model is the quotient space $\hat{X}=X/\Gamma$. The details of the construction will be explained in the next section. Here we merely mention the properties which are required to extract the relevant information about the four-dimensional theory. We first note that the freely-acting symmetry for our example is $\Gamma=\mathbb{Z}_2$, so that the symmetry order is $|\Gamma|=2$. For the number of Kahler moduli, $T^i$, we have~\footnote{The number of Kahler moduli is given by the Hodge number $h^{1,1}(\hat{X})$ of the quotient manifold. It turns out that for all models in the database this number equals $h^{1,1}(X)$, although this is not true in general.}
\begin{equation}
  \#(\mbox{Kahler moduli})=h^{1,1}(\hat{X})=h^{1,1}(X)
\end{equation}  
 The number of complex structure moduli, $Z$, is then given by
\begin{equation}
 \#(\mbox{complex structure moduli})=h^{2,1}(\hat{X})=h^{1,1}(X)-\frac{\eta(X)}{2|\Gamma|}\; .
\end{equation}
From Fig.~\ref{fig:cicy6732} we have $\eta(X)=-64$ and together with $h^{1,1}(X)=5$ and $|\Gamma|=2$ this implies that the model has $21$ complex structure moduli. Another relevant quantity which can be read off from Fig.~\ref{fig:cicy6732} is $\kappa$, defined in Eq.~\eqref{kappadef}, which determines the Kahler potential~\eqref{Kmod} for the Kahler moduli $T^i=t^i+2i\chi^i$. For our example it is given by
\bea \kappa &=& 12 t_1 t_2 t_3+ 12 t_1 t_2 t_4 + 12 t_1 t_3 t_4 +12
t_2 t_3 t_4 + 12 t_1 t_2 t_5 + 24 t_1 t_3 t_5 + 24 t_2 t_3 t_5 \nonumber\\
&&+ 24 t_1 t_4 t_5 +24 t_2 t_4 t_5 + 24 t_3 t_4 t_5 + 12 t_1
t_5^2 + 12 t_2 t_5^2+ 24 t_3 t_5^2 + 24 t_4 t_5^2 + 8 t_5^3 \;. \label{kappaex}
\eea
The database entry for our model which defines the vector bundle is shown in Fig.~\ref{fig:model7_1}.
As before, we defer the details of the construction to later and focus on how to extract the relevant low-energy quantities. The charge vector ${\bf k}_a$ which determine the $\anU$ transformations~\eqref{chitrafo} of the axions are given by the column vectors of the matrix $V$ in Fig.~\ref{fig:model7_1}. For our example this means
\begin{equation}
 ({\bf k}_1,\ldots ,{\bf k}_5)=\left(\begin{array}{rrrrr} 1&1&0&0&-2\\1&1&0&-2&0\\1&-1&-2&1&1\\0&1&1&-1&-1\\-1&-1&0&1&1\end{array}\right)\; . \label{ks}
\end{equation} 
The transformation of the dilatonic axion, $\sigma$, in Eq.~\eqref{chitrafo} also depends on the numbers $\beta_i$ which enter the gauge kinetic function~\eqref{f}. They can be computed from
\begin{equation}
 \beta_i=\frac{1}{|\Gamma|}\left(c_{2i}(V)-\frac{1}{2}c_{2i}(TX)\right)\; .
\end{equation}
From Fig.~\ref{fig:cicy6732} we read off $c_2(TX)=(24,24,24,24,56)$ and from Fig.~\eqref{fig:model7_1} we have $c_2(V)=(12,12,12,12,32)$. With $|\Gamma|=2$ this means that for our example
\begin{equation}
 {\boldsymbol \beta}=(-2,-2,0,2,2)\; . \label{betaex}
\end{equation} 
\subsubsection{The matter field sector}
All of the models in the database have a massless spectrum which
includes the gauge and matter spectrum of the MSSM. However, some models have 
additional massless fields at the Abelian locus, $S^\alpha=0$, which can include additional vector-like pairs of Higgs doublets
and one additional massless $U(1)$ gauge boson, related to the non-anomalous part of the $\anU$ gauge symmetry.
Masses for these fields may be generated by non-vanishing $S^\alpha$ VEVs and for this reason such models have been
included in the database.

Our example model has one additional massless $U(1)$ vector field as stated at the top of Fig.~\ref{fig:model7_1}. Alternatively, this follows from the general result~\eqref{numU1} and the fact that only three of the vectors ${\bf k}_a$ in Eq.~\eqref{ks} are linearly independent. The matter field spectrum at the Abelian locus, $S^\alpha=0$, can be read off from the database entry entitled ``Downstairs spectrum" and, from Fig.~\ref{fig:model7_1}, for our example model it is given by
\be
{\bf 10}_3,{\bf 10}_4,{\bf 10}_5,\bar{\bf 5}_{1,2},2
  \bar{\bf 5}_{4,5},H_{3,4},\bar{H}_{3,4},H_{3,5},\bar{H}_{3,5},S_{2,1},4
  S_{4,2},4 S_{5,2},2 S_{4,3},2 S_{5,3},S_{4,5},S_{5,4} \;.
\label{downspect}
\ee
Here, we follow the notation introduced in the previous section. In particular, we have grouped the standard model particles into their
standard $SU(5)$ representations for ease of notation. We recall that the subscripts indicate the $\anU$ charge of a multiplet. For example, ${\bf 10}_3$ denotes a ${\bf 10}$ multiplet of $SU(5)$ with $\anU$ charge ${\bf Q}({\bf 10}_3)={\bf e}_3$, while $\bar{\bf 5}_{1,2}$ denotes a $\bar{\bf 5}$ multiplet of $SU(5)$ with $\anU$ charge ${\bf Q}(\bar{\bf 5}_{1,2})={\bf e}_1+{\bf e}_2$. The $\anU$ charge for a down Higgs is, for example, ${\bf Q}(H_{3,4})={\bf e}_3+{\bf e}_4$ while we have ${\bf Q}(\bar{H}_{3,4})=-{\bf e}_3-{\bf e}_4$ for an up Higgs. The standard model singlet fields are denoted by $S$ and their $\anU$ charge pattern is exemplified by ${\bf Q}(S_{2,1})={\bf e}_2-{\bf e}_1$. 

The mixed ${\cal J}G_{\rm SM}G_{\rm SM}$ triangle anomaly can be computed from Eq.~\eqref{trianom}. For the above spectrum we easily find 
\begin{equation}
 {\bf A}=(1,1,3,5,5)\; .
\end{equation} 
Further, using the values of the charge vectors~\eqref{ks} and of ${\boldsymbol\beta}$ in Eq.~\eqref{betaex} it follows that
\begin{equation}
 (k^i_a\beta_i)_a=(-2,-2,0,2,2)\; . \label{kbetaex}
\end{equation}
Therefore, the anomaly constraint~\eqref{anomcons} is indeed satisfied for our example model, as it must be due to the Green-Schwarz mechanism. This simple calculation provides a useful consistency check for our models.

The spectrum~\eqref{downspect} shows that the example model contains two massless pairs of Higgs doublets at the locus $S^\alpha=0$. 
In cases such as these a physical pair of Higgs doublets is chosen and separate models are generated for each possible choice. For the case at hand,
the choice is $H_{3,5}$, $\bar{H}_{3,4}$, as the ``Phys.~Higgs" entry in Fig.~\ref{fig:model7_1} indicates. For consistency, the other Higgs doublet
should then obtain a mass from non-zero singlet VEVs if we are to recover exactly the standard model charged spectrum with the chosen Higgs doublet. The relevant mass operators will be discussed in Section \ref{sec:datahiggssector}. Another possibility is, of course, to consider phenomenological models with two or three Higgs doublets. 
\begin{figure}
\centering
\includegraphics[width=17.6cm]{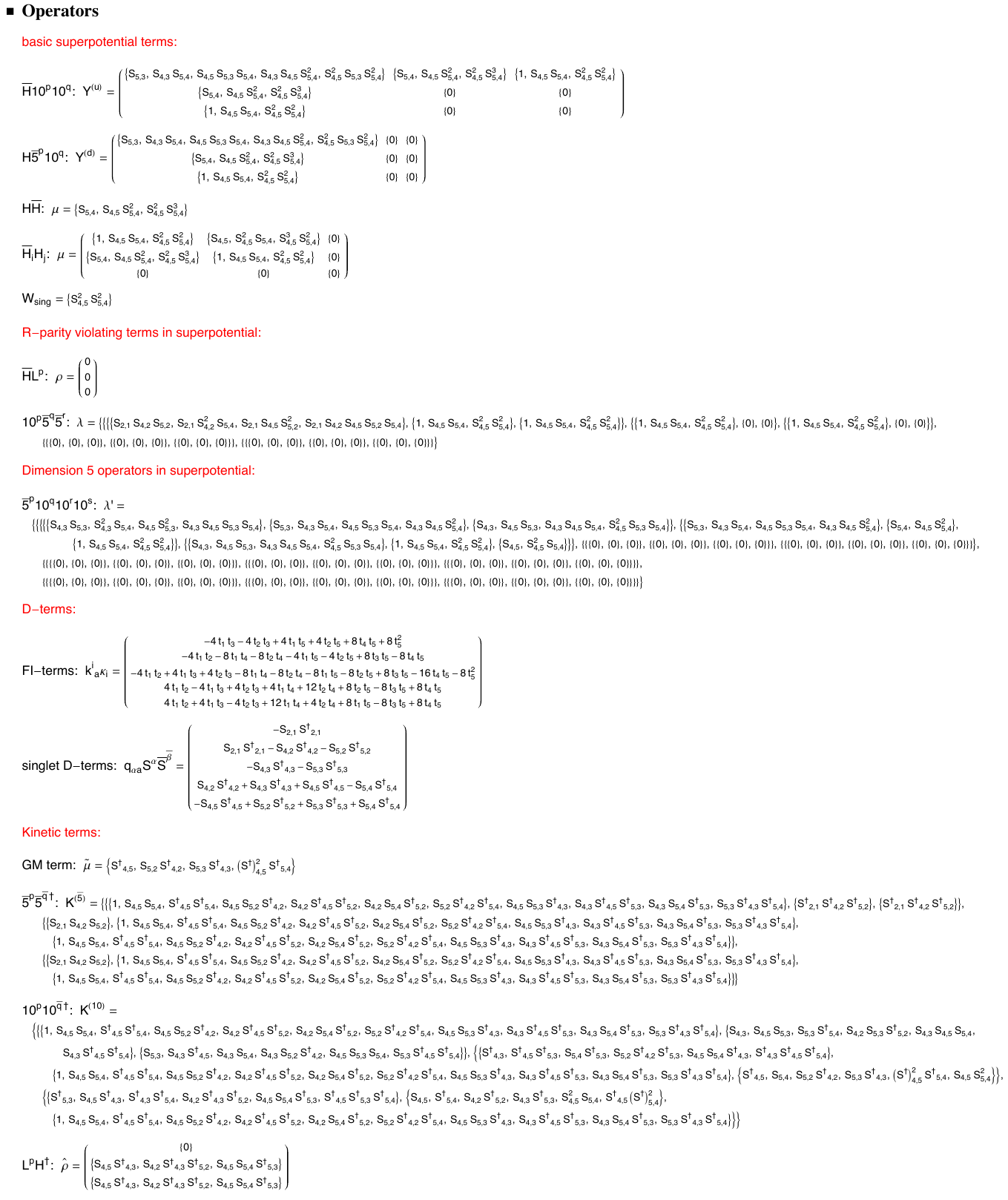}
\parbox{7in}{\caption{\it\small Example database entry specifying operators for the line bundle standard model in Fig.~\ref{fig:model7_1}.}\label{fig:model7_2}}
\end{figure}
In the following, we will discuss the various types of $\anU$ invariant operators allowed in the effective action and their possible phenomenological relevance. For our example model, the relevant database entry listing these operators is shown in Fig.~\ref{fig:model7_2}. 
\subsubsection{The D-terms}
\label{sec:datadterms}
The general form of the D-term has been given in Eq.~\eqref{Dterm}. The first term in this expression is the leading part of the FI term. Since we have already specified $\kappa$, for our example model given by Eq.~\eqref{kappaex}, it suffices to provide the expressions $k^i_a\kappa_i$ in order to fix this first term.  In our database, these expressions are listed under the heading ``FI-terms" and, from Fig.~\ref{fig:model7_2}, for our example are given by
\begin{equation}
(k^i_a\kappa_i)=
\left(
\begin{array}{l}
 -4 t_1 t_3-4 t_2 t_3+4 t_1 t_5+4 t_2 t_5+8 t_4 t_5+8 t_5^2 \\
 -4 t_1 t_2-8 t_1 t_4-8 t_2 t_4-4 t_1 t_5-4 t_2 t_5+8 t_3 t_5-8 t_4 t_5 \\
 -4 t_1 t_2+4 t_1 t_3+4 t_2 t_3-8 t_1 t_4-8 t_2 t_4-8 t_1 t_5-8 t_2 t_5+8 t_3 t_5-16 t_4 t_5-8 t_5^2 \\
 4 t_1 t_2-4 t_1 t_3+4 t_2 t_3+4 t_1 t_4+12 t_2 t_4+8 t_2 t_5-8 t_3 t_5+8 t_4 t_5 \\
 4 t_1 t_2+4 t_1 t_3-4 t_2 t_3+12 t_1 t_4+4 t_2 t_4+8 t_1 t_5-8 t_3 t_5+8 t_4 t_5
\end{array}
\right)\; .
\label{datafiterms}
\end{equation}
In order to specify the dilaton-dependent correction to the FI-term, which corresponds to the second term in Eq.~\eqref{Dterm}, we need to provide the vector $(k^i_a\beta_i)_a$. For the example model, this vector has already been determined from other database entries and is given in Eq.~\eqref{kbetaex}.
Finally, we need the last term in Eq.~\eqref{Dterm} which represents the matter field contribution. Here, we write down the parts of these matter field D-terms which depend only on the singlet fields $S^\alpha$. In the database it is listed under the heading ``singlet D-terms". Fig.~\ref{fig:model7_2} shows that for our example model it is given by
\be
\left(\sum_{\alpha,\bar{\beta}}q_{a\alpha}S^\alpha\bar{S}^{\bar{\beta}}\right)= \left(\begin{array}{c}
 -S_{2,1} S^{\dagger }_{2,1} \\
 S_{2,1} S^{\dagger }_{2,1}-S_{4,2} S^{\dagger }_{4,2}-S_{5,2} S^{\dagger }_{5,2} \\
 -S_{4,3} S^{\dagger }_{4,3}-S_{5,3} S^{\dagger }_{5,3} \\
 S_{4,2} S^{\dagger }_{4,2}+S_{4,3} S^{\dagger }_{4,3}+S_{4,5} S^{\dagger }_{4,5}-S_{5,4} S^{\dagger }_{5,4} \\
 -S_{4,5} S^{\dagger }_{4,5}+S_{5,2} S^{\dagger }_{5,2}+S_{5,3} S^{\dagger }_{5,3}+S_{5,4} S^{\dagger }_{5,4}
\end{array}
\right) \;.
\label{datadterms}
\ee
In writing these expressions, we have omitted the Kahler moduli space metric of the singlet fields which should appear but is not explicitly known. However, for the analysis of D-flat directions with $S^\alpha\neq 0$ is it usually sufficient to know that this metric is positive definite.

At the locus $S^\alpha=0$ where all singlet VEVs vanish, the systems we consider in this paper admit a solution to the D-term equations. A necessary condition for this to be possible is that the basic constraint~\eqref{numcons} holds, that is, that we have more Kahler moduli than linearly independent charge vectors ${\bf k}_a$. For our example model, which has five Kahler moduli and three linearly independent charge vectors, this is certainly satisfied. Note also that, since the overall scaling of the K\"ahler moduli does not enter the D-term equations, such a solution can always be scaled to the large volume regime (that is, the regime where all values of the Kahler moduli $t^i$ are large).

For non-vanishing singlet VEVs, $S^\alpha\neq 0$, the existence of D-flat directions depends on the details of the above singlet matter field terms in the D-term and has to be analyzed case by case. Of course, for supersymmetric vacua with $S^\alpha\neq 0$ we also need to check the F-term equations which follow from the singlet superpotential $W_{\rm sing}$ in Eq.~\eqref{W}. We now turn to a discussion of this singlet superpotential.
\subsubsection{The singlet superpotential: F-terms and Neutrino Majorana masses}
\label{sec:datafterms}
In the database, the singlet superpotential is denoted by $W_{\rm sing}$. A quick glance at Fig.~\ref{fig:model7_2} shows that for the example model it 
is given as
\be
W_{\rm sing} \sim S_{4,5}^2 S_{5,4}^2 \;,
\label{datasuperpot}
\ee
with possible higher dimension operators omitted. It is important to note that the singlet fields with a given $\anU$ charge can appear with multiplicity greater than one, as is evident from the spectrum~\eqref{downspect}. For simplicity, the sum over these multiplicities has been suppressed in the above expression.
We also re-iterate from our general discussion in the previous section, that $\anU$-invariance of singlet operators does not necessarily imply their presence in $W_{\rm sing}$, as they might be forbidden for other reasons. An example of this is provided by the gauge-invariant quadratic
terms in the singlets, $S_{4,5}S_{5,4}$ for our example model. We know these terms must vanish since the underlying string construction shows that all of the singlets are indeed massless at the locus $S^\alpha=0$. Given the uncertainty in the exact coefficients of the terms it is not possible to give an explicit
solution to the F-terms where the contribution of one operator cancels against another. However, one can argue for the existence of
such a solution assuming generic coefficients. It is also possible to show that, for a given combination of singlet VEVs, the contribution of each operator to the F-terms vanishes separately. For the above example (\ref{datasuperpot}) it is clear that the (global) singlet F-terms vanish as long as the VEVs of either $S_{4,5}$ or $S_{5,4}$ are zero. 

The standard model singlet fields are also attractive candidates for
right-handed neutrinos (RHNs). In this context, the role of the singlet
superpotential is to generate Majorana masses for the RHNs due to non-vanishing singlet VEVs .
For example, for the superpotential~\eqref{datasuperpot} a non-zero VEV for $S_{5,4}$ (with the VEV of
$S_{4,5}$ still vanishing to satisfy the F-term equations) generates a Majorana mass term for $S_{4,5}$ which might then play the role of a RHN.
Of course a realisation of the see-saw mechanism also requires the presence of an associated Dirac mass. This will be discussed in Section~\ref{sec:datarpar}. 

\subsubsection{The Higgs sector}
\label{sec:datahiggssector}

The only part of the spectrum charged under the standard model that
varies within the database is the Higgs sector - some examples contain
more than one set of Higgs doublets. For such models we identify a
particular pair of weak doublets to play the role of the Higgs fields.
This pair will then be used to calculate all of the relevant phenomenological operators
such as the Yukawa couplings. The remaining doublets will be considered as
exotic fields which must obtain a large mass. Another option would be
to consider theories with multiple light Higgs doublet pairs. To cover all
possibilities, we have generated a separate database entry for each possible choice of a Higgs doublet pair among the available doublets.
The most straightforward way for the additional doublets to obtain a mass is through couplings of the form $\mu_{ij}(S)H^i\bar{H}^j$.
To study these couplings, for models with multiple doublet pairs, the database contains 
the $3\times 3$ mass matrix $\mu_{ij}$ for up to 3 pairs of Higgs fields. For our example model in Fig.~\ref{fig:model7_2} this matrix is given by
\be
\mu_{ij}= \left(\begin{array}{ccc}
 \left\{1,S_{4,5} S_{5,4},S_{4,5}^2 S_{5,4}^2\right\} & \left\{S_{4,5},S_{4,5}^2 S_{5,4},S_{4,5}^3 S_{5,4}^2\right\} & \{0\} \\
 \left\{S_{5,4},S_{4,5} S_{5,4}^2,S_{4,5}^2 S_{5,4}^3\right\} & \left\{1,S_{4,5} S_{5,4},S_{4,5}^2 S_{5,4}^2\right\} & \{0\} \\
 \{0\} & \{0\} & \{0\}
\end{array}
\right)\; .
\ee
For the case at hand one row and one column vanish because the model only contains two doublet pairs.
Our notation is such that each matrix entry $(i,j)$ lists singlet operator which can couple to $H^i\bar{H}^j$ in a $\anU$-invariant way.
The indices $i,j$ run over the massless doublet pairs in the order in which they are given in the
spectrum (\ref{downspect}). For a given singlet VEVs it is possible to
study if the additional doublets can indeed obtain a large mass while keeping the
chosen Higgs pair light. The diagonal entries in the above matrix contain entries $1$, consistent with the $\anU$-invariance
of the operator $H^i\bar{H}^i$. However, these entries should be ignored since, by construction, all doublets are indeed exactly massless 
at the locus $S^\alpha=0$. This is another example of a set of operators absent for reasons unrelated to $\anU$-invariance. 

Keeping models with multiple Higgs pairs in the database (while we did
not keep models with additional massless matter which could similarly
be made massive by GUT singlets) is primarily motivated by the
possibility of realising an approximate $U_{\rm PQ}(1)$ symmetry. As will
be discussed in detail in section \ref{genform}, for the simplest
cases which we have scanned over, a single Higgs pair is always
vector-like under $\anU$ so that the $\anU$-symmetry does not contain a $U(1)_{PQ}$ symmetry.
However, such a symmetry may may be useful for forbidding
proton decay operators, for generating attractive flavour structures,
and for additional control over the Higgs mass. Given two pairs
of doublets, where each pair has different $\anU$ charges, it is possible to identify an off-diagonal combination as the
Higgs fields thereby inducing a $U(1)_{PQ}$. Indeed, this has been done for the example model in Fig.~\ref{fig:model7_1} where
$H_{3,5}$, $\bar{H}_{3,4}$ has been chosen as the physical Higgs pair. Of course the remaining doublets, $\bar{H}_{3,5}$ and $H_{3,4}$ in the example, must obtain a mass due which breaks the $U(1)_{PQ}$ symmetry.
If this breaking is sufficiently controlled, for example due to a breaking scale well below the string scale,
the remaining approximate $U(1)_{PQ}$ symmetry might still be useful.

For the example model, a VEV for $S_{4,5}$ gives mass to the additional doublets while keeping the physical Higgs fields massless.

\subsubsection{Yukawa Couplings}
\label{sec:datayukawas}

The database contains the Yukawa couplings under the headings $Y^{(u)}$ and $Y^{(d)}$. For the example model in Fig.~\ref{fig:model7_2} they are given by
\bea
 Y^{(u)}&=&\left(
\begin{array}{ccc}
 \left\{S_{5,3},S_{4,3} S_{5,4},S_{4,5} S_{5,3} S_{5,4}\right\} & \left\{S_{5,4},S_{4,5} S_{5,4}^2\right\} & \left\{1,S_{4,5} S_{5,4}\right\} \\
 \left\{S_{5,4},S_{4,5} S_{5,4}^2\right\} & \{0\} & \{0\} \\
 \left\{1,S_{4,5} S_{5,4}\right\} & \{0\} & \{0\}
\end{array}
\right) \nn \\
Y^{(d)}&=& \left(
\begin{array}{ccc}
 \left\{S_{5,3},S_{4,3} S_{5,4},S_{4,5} S_{5,3} S_{5,4}\right\} & \{0\} & \{0\} \\
 \left\{S_{5,4},S_{4,5} S_{5,4}^2\right\} & \{0\} & \{0\} \\
 \left\{1,S_{4,5} S_{5,4}\right\} & \{0\} & \{0\}
\end{array}
\right) \;,  \eea
where we have dropped terms higher than cubic in $S^\alpha$. For each model we also list the generic rank of these Yukawa matrices as in Fig.~\ref{fig:model7_1}. For the above matrices we have ${\rm rk}\left(Y^{(u)}\right)=\left\{2,2\right\}$ and ${\rm rk}\left(Y^{(d)}\right)=\left\{1,1\right\}$. Here, the first (second) entry denotes the generic rank if all $S^\alpha=0$ (if all $S^\alpha$ are non-zero). We note that a non-vanishing top Yukawa coupling of order one is possible in these models, even at the Abelian locus $S^\alpha=0$.

\subsubsection{Proton decay}
\label{sec:datadim4proton}

Proton decay forms one of the classic constraints on extensions of the
standard model. Dimension four proton decay operators are tightly
constrained by experiments to have coefficients less that $10^{-6}$
for any combination of generation indices \cite{Barbier:2004ez}. Often these are
forbidden by imposing the R-parity of the MSSM. However, within
the context of top-down model building from string theory, such an R-parity does not necessarily have to be realized.
It is, therefore, important to consider whether such operators can be forbidden using the $U(1)$ symmetries of our models. The
dimension four and five proton decay operators in~\eqref{WR} and \eqref{Wprot5} are denoted by $\lambda_{pqr}$ and $\lambda'_{pqrs}$, respectively,
and are listed under those headings in the database. For our example model they can be found in Fig.~\ref{fig:model7_2}.

\subsubsection{Bilinear R-parity violating operators and superpotential Neutrino Dirac masses}
\label{sec:datarpar}

An important set of operators which relates directly to neutrino physics are operators of the type $\rho_p\bar{H} L^p$ which appear in the superpotential~\eqref{WR}. Such R-parity violating operators, among other things, lead directly to large neutrino masses through mixing with the Higgs fields and so are constrained to be very small
(around $10^{-22}$ in Planck units). For the example model in Fig.~\ref{fig:model7_2} these coefficients vanish
\be \rho
=\left(
\begin{array}{c}
 0 \\
 0 \\
 0
\end{array}
\right) \;.
\ee
Therefore, in this case non-vanishing singlet VEVs cannot generate any R-parity violating operators. As ever, a bare quadratic term in the
superpotential is forbidden by construction in these models. 

If we take some of the standard model singlets to be RHNs then these
same terms also play the role of the superpotential neutrino Dirac
masses. These may then be combined with the pure singlet terms
discussed in section \ref{sec:datafterms} to realize the see-saw mechanism.
For the example model no such superpotential Dirac masses are allowed.

\subsubsection{Neutrino K\"ahler potential Dirac masses}
\label{sec:dataneutkah}

In the absence of superpotential Dirac or Majorana neutrino masses
there is a natural way to induce neutrino masses of the correct magnitude
through a K\"ahler potential operator \cite{ArkaniHamed:2000bq}. The relevant operator
in the matter Kahler potential~\eqref{Kmatter} is
\begin{equation}
 \hat{\rho}_pL^pH^\dagger\; .
\end{equation} 
When the up-type Higgs develops a VEV, $v$, it induces an F-term for the down-type
Higgs which, from the above operator, leads to Dirac neutrino masses.  For our example model in Fig.~\ref{fig:model7_2}
these couplings are given by
\be
\hat{\rho}=\left(
\begin{array}{c}
 \{0\} \\
 \left\{S_{4,5} S^{\dagger }_{4,3},S_{4,2} S^{\dagger }_{4,3} S^{\dagger }_{5,2},S_{4,5} S_{5,4} S^{\dagger }_{5,3}\right\} \\
 \left\{S_{4,5} S^{\dagger }_{4,3},S_{4,2} S^{\dagger }_{4,3} S^{\dagger }_{5,2},S_{4,5} S_{5,4} S^{\dagger }_{5,3}\right\}
\end{array}
\right) \;.
\ee
Following the discussion in Section~\ref{sec:datafterms} we may consider $S_{4,5}$ as a RHN. Then giving a VEV to $S_{4,3}$ induces Dirac neutrino mass. Note that we allow for conjugates of the singlets to appear since we are dealing with a K\"ahler potential operator.

\subsubsection{The Giudice-Masiero term}
\label{sec:datagmterm}

A well-known way to induce a $\mu$-term within
gravity mediated supersymmetry breaking is through the Giudice-Masiero
mechanism \cite{Giudice:1988yz}. The relevant operator in the matter field Kahler potential~\eqref{Kmatter}
is
\begin{equation}
 \tilde{\mu}H\bar{H}\; .
\end{equation} 
If $\tilde{\mu}$ depends on the conjugate, $S^{\alpha\dagger}$, of a singlet field which breaks supersymmetry, a $\mu$-term of the right order of magnitude is generated. Hence, we should list all gauge invariant operators of the above form which involve at least one singlet appearing as a conjugate. For the example model in Fig.~\ref{fig:model7_2} this leads to the operators
\be
\tilde{\mu}\supset\left\{S^{\dagger}_{4,5},S_{5,2}
  S^{\dagger}_{4,2},S_{5,3}
  S^{\dagger}_{4,3},\left(S^{\dagger}_{4,5}\right)^2
  S^{\dagger}_{5,4}\right\} \; .
\ee

\subsubsection{Kinetic terms and soft masses}
\label{sec:datakinsoft}

One of the useful properties of the $U(1)$ symmetries is that they allow us to gain a handle on the form of the kinetic terms of the matter fields. The kinetic terms enter the determination of the physical Yukawa couplings from the Yukawa couplings in the superpotential and are therefore of great importance. However, due to their non-holomorphic nature, they are rather difficult to calculate from first principles. For our example model in Fig.~\ref{fig:model7_2} we have
\bea
K^{(\bar{\bf 5})}&=&\left(
\begin{array}{ccc}
 1 & S^{\dagger }_{2,1} S^{\dagger }_{4,2} S^{\dagger }_{5,2} & S^{\dagger }_{2,1} S^{\dagger }_{4,2} S^{\dagger }_{5,2} \\
 S_{2,1} S_{4,2} S_{5,2} & 1 & 1 \\
 S_{2,1} S_{4,2} S_{5,2} & 1 & 1 
\end{array}
\right) \;, \nn \\
K^{({\bf 10})}&=&\left(
\begin{array}{ccc}
 1 & S_{4,2} &  S_{5,3} \\
 S_{4,3}^{\dagger} & 1 & S_{4,5}^{\dagger} \\
 S_{5,3}^{\dagger} & S_{4,5} & 1 
\end{array}
\right) \;. \label{datakin}
\eea
For simplicity we have only displayed the leading term for each operator while Fig.~\ref{fig:model7_2} shows the full list including terms involving up to 3 standard model singlets.

The same gauge invariant combinations are also relevant for
constraining the possible soft supersymmetry breaking masses that can
appear in the potential. Understanding their flavour structure is
important, especially within gravity mediation, due to the possibility
of inducing flavour changing neutral currents (FCNCs). It is well
known that it is possible to use $U(1)$ global symmetries to constrain
the flavour off-diagonal terms of such soft masses and the structure of the operators
in Eq.~\eqref{datakin} will be conductive to realizing such a scenario.

\subsection{General phenomenology overview of database}
\label{sec:genoverdata}

Having described the output data for an individual model it is
interesting to consider how various basic phenomenological properties
are distributed within the database as a whole. To this end, we
present some statistics of models in the database. The given numbers are not meant
as a comprehensive statistical analysis of heterotic line bundle models but merely as
a rough indication of how difficult it might be to achieve certain phenomenological properties within this class.

The models presented in the database~\cite{database} descend from the 202 $SU(5)$ GUT models constructed in Ref.~\cite{Anderson:2011ns} by quotienting the Calabi-Yau three-fold and Wilson line breaking. This process breaks the GUT group to the standard model group and projects out certain unwanted states, in particular the Higgs triplets still present in the GUT theory. Depending on the symmetry by which we divide (either $\mathbb{Z}_2$ or $\mathbb{Z}_2\times\mathbb{Z}_2$ in all cases) there are between order 100 and 1000 choices per GUT model on how to realize this breaking. Not all of these choices lead to a phenomenologically viable spectrum (for example, in some cases Higgs triplets are still present) and, for a given model, many choices result in the same spectrum. In our scan, we have only kept the cases which lead to an acceptable spectrum and we have chosen one representative model per spectrum generated. This leads to a total of $2122$ line bundle standard models which originate from the 202 GUT models. A list of these models is available as a data file at~\cite{database}. Upon inspection it turns out that many of these models are closely related in that they have the same spectrum and are based on the same (or equivalent) Calabi-Yau manifolds and the same bundle. Two models related in this way look identical for the purposes of this paper, although, since they are generally based on different symmetries of the Calabi-Yau manifold, they may differ at a more detailed level. We have eliminated these redundancies in the explicit printout of the models, in order to keep the size manageable. This results in 407 models available in the printed lists at~\cite{database}. The statistics of phenomenological properties below is based on these 407 models.

The results are summarized in Table~\ref{tab:dbstat} below.
\begin{table}[!h]
\begin{center}
\begin{tabular}{|l|l|l|l|l|l|l|l|}
 \hline
 {\fn standard}&{\fn no mass-}&{\fn 1 Higgs}\fn &{\fn 2 Higgs}&{\fn 3 Higgs}&{\fn rk$(Y^{(u)})$}&{\fn no proton decay,}&{\fn 1 Higgs, rk$(Y^{(u)})>0$,}\\
{\fn  models}&{\fn less $U(1)$}&{\fn pair}&{\fn pairs}&{\fn pairs}&{\fn $>0$}&{\fn $\lambda=\lambda'=0$}&{\fn $\lambda=\lambda'=0$, $U(1)$s massive}\\\hline
{\fn 407}&{\fn 237}&{\fn 262}&{\fn 77}&{\fn 63}&{\fn 45}&{\fn 198}&{\fn 13}\\\hline
 \end{tabular}
 \parbox{7in}{\caption{\it\small Statistics of basic properties in the standard model database~\cite{database}.}\label{tab:dbstat}}
 \end{center}
 \end{table}
 A few comments on what precisely is being counted are in order. The number of massless $U(1)$ vector fields and the number of Higgs pairs is determined at the Abelian locus $S^\alpha=0$ where all singlet VEVs vanish. As discussed earlier, massless $U(1)$ vector bosons can acquire a mass when singlet VEVs are switched on. This means that the 170 models with such a massless vector boson are not necessarily ruled out but have to be analyzed in more detail. A similar remark applies to models with more than one Higgs pair. The rank of the up Yukawa matrix $Y^{(u)}$ in column six of the table has also been determined for vanishing singlet VEVs. It can be shown that the $U(1)$ symmetries in $\anU$ never allow an up Yukawa matrix with rank one and, it turns out there are no examples with ${\rm rk}(Y^{(u)})=3$ in our list. This means all $45$ models mentioned in column six have $Y^{(u)}=2$ while all remaining models have an entirely vanishing up Yukawa matrix for vanishing singlet VEVs. A positive rank for $Y^{(u)}$ is, of course, desirable since we would like a top Yukawa coupling of order one, however, it would be preferable to have ${\rm rk}(Y^{(u)})=1$. This can, in fact, be achieved for related constructions, to be discussed in the second part of the paper, which lead to fewer $U(1)$ symmetries in the low-energy theory. 

The second last column in the table gives the number of models for which all proton decay operators in~\eqref{WR} and \eqref{Wprot5} vanish, that is, $\lambda_{pqr}=0$ and $\lambda'_{pqrs}=0$ for all values of the family indices and in the presence of generic singlet VEVs. Evidently, this is a fairly strong condition which is sufficient but not necessary to guarantee that such operators do not destabilize the proton. For example, some terms for the second and third family might be allowed, particularly if they are suppressed by small singlet VEVs. This has to be studied in detail on a case-by-case basis.
At any rate, it is encouraging that we remain with $13$ models even when all conditions are imposed simultaneously, as in the last column of Table~\ref{tab:dbstat}.

\section{The Geometry of Split Heterotic Models} \label{mrsection4sir}
In this section, we introduce the necessary formalism to study compactifications of the $E_8\times E_8$ heterotic theory on smooth Calabi-Yau manifolds \cite{gsw} and the associated low-energy particle physics. In particular, we will discuss split bundles, that is, bundles with a direct product structure group. Bundles of this type, with the simplest splitting into a structure group $S(U(1)^5)$, underly the standard models presented in the first part of this paper. For reason which will become clear we will keep our discussion more general to cover all splittings into unitary factors. The construction of specific models based on this formalism will be presented in the next section.

\subsection{General Formalism}
\label{genform}
We begin by briefly reviewing the structure and constraints of generic heterotic Calabi-Yau compactifications. 
The geometric data required to specify a heterotic Calabi-Yau compactification which preserves four-dimensional $N=1$ supersymmetry consist of a Calabi-Yau three-fold, $X$, two holomorphic, poly-stable vector bundles, $V$ and $\tilde{V}$, with zero slope over $X$ and a holomorphic curve $C$ with second homology class $[C]$. The two vector bundles are associated to the observable and hidden $E_8$ sectors of the theory and their structure groups, $H$ and $\tilde{H}$, must be sub-groups of $E_8$. In the present paper we will take these structure groups to be $SU(n)$ (hence $c_1(V)=0$), typically with $n=5$ for the observable sector, or sub-groups thereof. The holomorphic curve $C$ is wrapped by five-branes (NS five-branes in the weakly coupled limit, M five-branes in the 11-dimensional strong-coupling picture) whose other directions stretch across the four-dimensional uncompactified space-time. 

This data has to satisfy a series of consistency conditions in order to obtain a well-defined vacuum. We will outline the conditions briefly here and study them in more depth in the following subsections. The first condition on the geometry is the well-known heterotic anomaly cancellation condition \cite{gsw},
\begin{equation}
 c_2(TX)-c_2(V)-c_2(\tilde{V})=[C] \label{ancond}\; .
\end{equation}
In the subsequent discussion, we will focus on the observable bundle $V$. The hidden bundle $\tilde{V}$ and the five-brane curve $C$ will not be constructed explicitly but we will ensure, by an appropriate choice of $V$, that a consistent completion of the model exists. Usually, we will do this by requiring $c_2(TX)-c_2(V)$ to be an effective class, $[C] \in H_2(X)$. Hence, in this case we can obtain a consistent completion of the model by adding an appropriate amount of five-branes while choosing the hidden bundle $\tilde{V}$ to be trivial.

The presence of the vector bundle $V$ (that is, the presence of non-trivial gauge field VEVs over the Calabi-Yau three-fold, $X$) breaks the visible sector $E_8$ symmetry to a sub-group, $G \subset E_8$. The gauge group, $G$, is given by the commutant of $H$ in $E_8$. For example,  choosing the structure group to be $H=SU(5)$ produces the commutant of $G=SU(5)$ so that we obtain a minimal GUT theory in four dimensions. If $H$ is a proper rank four sub-group of $SU(5)$, as we will consider in this paper, the low-energy gauge group enhances to $G=SU(5)\times \anU$, where $\anU\cong U(1)^{f-1}$ consists of a product of $U(1)$ factors. As will be reviewed in the following sections, it is well-known, that some or all of these $U(1)$ factors are anomalous in the Green-Schwarz sense and are, consequently, spontaneously broken at a high scale with associated massive vector bosons. The various types of low-energy multiplets in the GUT theory are obtained by decomposing the ${\bf 248}$ adjoint representation of $E_8$ into representations of $H \times G$ and the number of each multiplet can be computed from the bundle-valued cohomology of $V$ and its tensor powers. 

In order to produce realistic four-dimensional models, it is necessary to further break the GUT group to the Standard Model. To this end, we will also introduce Wilson lines. However, Wilson lines can only be defined over a Calabi-Yau manifold, $X$, which is not simply connected (i.e. $\pi_1(X) \neq 0$). Since there are few known Calabi-Yau geometries which have a non-trivial fundamental group by construction, we shall explicitly construct such manifolds from simply connected ones, by forming quotient manifolds $X/\Gamma$ where $\Gamma$ is a discrete group. To this end, we require the existence of a symmetry $\Gamma$ acting freely on the Calabi-Yau three-fold $X$, so that the quotient $\hat{X}=X/\Gamma$ is smooth and has a non-trivial first fundamental group (for example, for $X/\mathbb{Z}_n$, $\pi_1(X)=\mathbb{Z}_n$). In order for the bundle, $V$, to descend to a bundle, $\hat{V}$, on the quotient Calabi-Yau $\hat{X}$ the group $\Gamma$ must act consistently on the bundle. This means, there must be a group action of $\Gamma$ on $V$ which commutes with the projection $\pi: V \to X$ and satisfies a certain co-cycle condition. Such a group action is referred to as an  ``equivariant structure" and a bundle that admits such a structure is called ``equivariant" with respect to $\Gamma$. In summary, the ``downstairs" Calabi-Yau manifold $\hat{X}$ is defined by a multi-sheeted cover $q: X \to X/\Gamma$ and all vector bundles on $\hat{X}$ can be pulled back to equivariant bundles $V=q^* {\hat V}$ on $X$. That is, if $V$ is equivariant, ${\hat V}$ is well-defined on ${\hat X}$.

With the addition of non-trivial Wilson lines, the full ``downstairs" bundle on $\hat{X}$ is $\hat{V}\oplus{\cal  W}$, where ${\cal W}$ is a flat rank one bundle representing an Abelian Wilson line. Its structure group can be embedded into hypercharge $U_Y(1)\subset SU(5)$ in order to break $SU(5)$ into the standard model group. The downstairs zero-mode spectrum can be obtained from the bundle cohomology of $\hat{V}\oplus {\cal W}$ which, in practice, can be computed from the cohomology of the upstairs bundle $V$ and its equivariant structure.  

With this framework in hand, we turn now to the central point of this paper. We will consider vector bundles $V$ with a split structure group of the form
\begin{equation}
 H=S(U(n_1)\times\dots\times U(n_f))\; ,
\end{equation} 
where $n_a\geq1$ are integers. In order to ensure that the non-Abelian part of the low-energy gauge group is given by the GUT group $SU(5)$ we will demand that  $\sum_{a=1}^fn_a=5$. We will be especially interested in the case of ``maximal splitting" when $n_a=1$ for all $a$. In this case, $V$ is simply a direct sum of five line bundles, and $SU(5)$ splits as $H=S(U(1)^5)$. However, other patterns will be of interest as well so that we keep the formalism general for now and characterize a particular pattern by the integer vector ${\bf n}=(n_1,\ldots ,n_f)$. We will now explain in detail how the general formalism for heterotic Calabi-Yau compactifications outlined above applies to such split bundles. We begin with some simple group theoretical considerations. 

\subsection{Group theory}\label{gen_split}
In this section we lay down the necessary notation to discuss both the gauge symmetries of the four-dimensional theory, as well as the structure group of the visible sector bundle, $V$.
We will find it convenient to introduce two sub-groups of the bundle structure group $H$, namely the maximal semi-simple subgroup $H_s$ and the maximal Abelian sub-group $\anU$. Explicitly, they are given by
\begin{equation}
 H_s=SU(n_1)\times\dots\times SU(n_f)\; ,\quad \anU\cong\{(e^{i\eta^1},\ldots,e^{i\eta^{f}})|\sum_{a}n_a\eta^a=0\}\cong U(1)^{f-1}\; ,
 \label{HsJ}
\end{equation} 
where $\eta^a$ are the $U(1)$ group parameters and the sum condition in the definition of $\anU$ accounts for the fact that $H$ consists of special unitary matrices. The different possible splittings of $SU(5)$, together with the sub-groups $H_s$ and $\anU$ are listed in Table~\ref{tab1}.
\begin{table}[h]
\begin{center}
\begin{tabular}{|l|l|l|l|l|}\hline
$H$&$H_s$&$\anU$&$f$&${\bf n}$\\\hline\hline
$S(U(4)\times U(1))$&$SU(4)$&$U(1)$&$2$&$(4,1)$\\\hline
$S(U(3)\times U(2))$&$SU(3)\times SU(2)$&$U(1)$&$2$&$(3,2)$\\\hline
$S(U(3)\times U(1)^2)$&$SU(3)$&$U(1)^2$&$3$&$(3,1,1)$\\\hline
$S(U(2)^2\times U(1))$&$SU(2)^2$&$U(1)^2$&$3$&$(2,2,1)$\\\hline
$S(U(2)\times U(1)^3)$&$SU(2)$&$U(1)^3$&$4$&$(2,1,1,1)$\\\hline
$S(U(1)^5)$&$1$&$U(1)^4$&$5$&$(1,1,1,1,1)$\\\hline
\end{tabular}
\parbox{6in}{\caption{\it\small The six splittings $H$ of $SU(5)$ considered in this paper, together with the maximal semi-simple sub-groups $H_s\subset H$, the maximal Abelian sub-groups $\anU\subset H$, the number $f$ of factors and the split vector ${\bf n}$.}\label{tab1}}
\end{center}
\end{table}
For the subsequent discussion it will be useful to label representations of the group $H$ by the $H_s$ and $\anU$ representations they induce. We denote by ${\cal F}_a$ (${\rm Adj}_a$) the representation of $H_s$ which transforms as a fundamental (adjoint) of the $SU(n_a)$ factor in $H_s$ and as a singlet under all other factors. Representations of $\anU$ are specified by a charge vector ${\bf q}=(q_1,\ldots ,q_f)$. As a consequence of the constraint in the definition~\eqref{HsJ} of $\anU$, in order to get a one-to-one correspondence between charge vectors and $\anU$ representations, we have to identify two such vectors ${\bf q}$ and $\tilde{\bf q}$ if
\begin{equation}
 {\bf q}-\tilde{\bf q}\in\mathbb{Z}{\bf n}\; . \label{qid1}
\end{equation}
Finally, an $H$ representation which transforms under the representation $R$ of $H_s$ and carries $\anU$ charge ${\bf q}$ is denoted by $R_{\bf q}$.  Using this notation we can write down rules for the branching of $SU(5)$ representations into $H$ representations. For the $SU(5)$ representations relevant to our discussion these branching rules read explicitly

\begin{equation}
\begin{array}{lllllll}
 {\bf 10}&\rightarrow&(\bigoplus_{a=1}^f(\wedge^2{\cal F}_a)_{2{\bf e}_a})\oplus (\bigoplus_{a<b}({\cal F}_a\otimes {\cal F}_b)_{{\bf e}_a+{\bf e}_b}) &&\bar{\bf 5}&\rightarrow&\bigoplus_{a=1}^f(\bar{{\cal F}}_a)_{-{\bf e}_a}\\
\bar{\bf 10}&\rightarrow&(\bigoplus_{a=1}^f(\wedge^2\bar{{\cal F}}_a)_{-2{\bf e}_a})\oplus(\bigoplus_{a<b}(\bar{{\cal F}}_a\otimes \bar{{\cal F}}_b)_{-{\bf e}_a-{\bf e}_b})&&{\bf 5}&\rightarrow&\bigoplus_{a=1}^f({\cal F}_a)_{{\bf e}_a}\\
 {\bf 24}&\rightarrow&\bigoplus_a({\rm Adj}_a)_{{\bf 0}}\oplus\bigoplus_{a\neq b}({\cal F}_a\otimes\bar{{\cal F}}_b)_{{\bf e}_a-{\bf e}_b}&&&&
\end{array} 
\label{SU5branching}
\end{equation}
where ${\bf e}_a$ denotes the $a^{\rm th}$ standard unit vector. 

Let us now embed $H$ into $E_8$ via the embedding chain $H\subset SU(5)\subset E_8$. The commutant of the so-embedded $H$ within $E_8$, that is the low-energy gauge group, is given by $G=SU(5)\times \anU$. As discussed before, the $U(1)$ factors in $\anU$ may be Green-Schwarz anomalous in which case their associated gauge bosons are massive. In order to find the multiplet types in the resulting $SU(5)$ GUT theory we need to decompose the ${\bf 248}$ adjoint representation of $E_8$. We begin with its well-known branching under $SU(5)\times SU(5)$ given by
\begin{equation}
 {\bf 248}\rightarrow ({\bf 1},{\bf 24})\oplus({\bf 24},{\bf 1})\oplus({\bf 10},\bar{\bf 5})\oplus({\bf 5},{\bf 10})\oplus(\bar{\bf 10},{\bf 5})\oplus(\bar{\bf 10},\bar{\bf 5})\; .
\end{equation} 
Here, we think of the first $SU(5)$ as the internal and the second $SU(5)$ as the external gauge group.
If we replace the internal $SU(5)$ representations with the branching rules in~\eqref{SU5branching} we immediately obtain the desired branching of ${\bf 248}$ into representations of $H\times SU(5)$. The resulting multiplets together with other relevant information are listed in Table~\ref{tab2}.
\begin{table}[h]
\begin{center}
\begin{tabular}{|l|l|l|l|l|l|}\hline
{\fn $H\cong H_s\times \anU$ repr. $R_{\bf q}$}&{\fn $SU(5)\times \anU$}&{\fn ass.~bundle}&{\fn $G_{\rm SM}\times \anU$ repr.}&{\fn symbol}&{\fn name}\\
&{\fn repr. $r_{\bf q}$}&{\fn $U_{R_{\bf q}}$}&&&\\\hline\hline
${\rm Adj}_{\bf 0}$&${\bf 1}_{\bf 0}$&$U_a\otimes U_a^*$&$(1,1)_{0,{\bf 0}}$&$S_a$&{\fn bundle modulus}\\\hline
$({\cal F}_a\otimes \bar{{\cal F}}_b)_{{\bf e}_a-{\bf e}_b}$, {\fn $a\neq b$}&${\bf 1}_{{\bf e}_a-{\bf e}_b}$&$U_a\otimes U_b^*$&$(1,1)_{0, {{\bf e}_a-{\bf e}_b}}$&$S_{ab}$&{\fn bundle modulus}\\\hline
$(\wedge^2 {\cal F}_a)_{2{\bf e}_a}$&$\bar{\bf 5}_{2{\bf e}_a}$&$\wedge^2U_a$&$(\bar{3},1)_{2,2{\bf e}_a}$&$d_a,T_a$&{\fn RH d quark/Higgs triplet}\\
&&&$(1,2)_{-3,2{\bf e}_a}$&$L_a,H_a$&{\fn LH lepton/d Higgs}\\\hline
$({\cal F}_a\otimes {\cal F}_b)_{{\bf e}_a+{\bf e}_b}$&$\bar{\bf 5}_{{\bf e}_a+{\bf e}_b}$&$U_a\otimes U_b$&$(\bar{3},1)_{2,{\bf e}_a+{\bf e}_b}$&$d_{ab},T_{ab}$&{\fn RH d quark/Higgs triplet}\\
\;\;{\fn $a<b$}&&&$(1,2)_{-3,{\bf e}_a+{\bf e}_b}$&$L_{ab},H_{ab}$&{\fn LH lepton/d Higgs}\\\hline
$({\cal F}_a)_{{\bf e}_a}$&${\bf 10}_{{\bf e}_a}$&$U_a$&$(1,1)_{6,{\bf e}_a}$&$e_a$&{\fn RH electron}\\
&&&$(\bar{3},1)_{-4,{\bf e}_a}$&$u_a$&{\fn RH u quark}\\
&&&$(3,2)_{1,{\bf e}_a}$&$Q_a$&{\fn LH quarks}\\\hline
$(\wedge^2 \bar{{\cal F}}_a)_{-2{\bf e}_a}$&${\bf 5}_{-2{\bf e}_a}$&$\wedge^2U_a^*$&$(3,1)_{-2,-2{\bf e}_a}$&$\tilde{d}_a,\bar{T}_a$&{\fn RH mirror d/Higgs triplet}\\
&&&$(1,2)_{3,-2{\bf e}_a}$&$\tilde{L}_a,\bar{H}_a$&{\fn LH mirror lepton/u Higgs}\\\hline
$(\bar{{\cal F}}_a\otimes \bar{{\cal F}}_b)_{-{\bf e}_a-{\bf e}_b}$&${\bf 5}_{-{\bf e}_a-{\bf e}_b}$&$U_a^*\otimes U_b^*$&$(3,1)_{-2,-{\bf e}_a-{\bf e}_b}$&$\tilde{d}_{ab},\bar{T}_{ab}$&{\fn RH mirror d/Higgs triplet}\\
\;\;{\fn $a<b$}&&&$(1,2)_{3,-{\bf e}_a-{\bf e}_b}$&$\tilde{L}_{ab},\bar{H}_{ab}$&{\fn LH mirror lepton/u Higgs}\\\hline
$(\bar{{\cal F}}_a)_{-{\bf e}_a}$&$\bar{\bf 10}_{-{\bf e}_a}$&$U_a^*$&$(1,1)_{-6,-{\bf e}_a}$&$\tilde{e}_a$&{\fn RH mirror electron}\\
&&&$(3,1)_{4,-{\bf e}_a}$&$\tilde{u}_a$&{\fn RH mirror u quark}\\
&&&$(\bar{3},2)_{-1,-{\bf e}_a}$&$\tilde{Q}_a$&{\fn LH mirror quark}\\\hline
\end{tabular}
\parbox{6.9in}{\caption{\it\small Representation content of the ${\bf 248}$ adjoint of $E_8$ under the subgroup $H\times SU(5)$. The first column provides the representation of the bundle structure group H in terms of $H_s\times \anU$, the second column the representation under the low-energy GUT gauge group $SU(5)\times \anU$, with the subscripts indicating the $\anU$ charge. The bundle associated to each representation is given in column three. Column four provides the break-up into representations of standard model group $G_{\rm SM}=SU_c(3)\times SU_W(2)\times U_Y(1) \times \anU$, with the first subscripts denoting the charge $3Y$ and the second subscripts the $\anU$ charge.}\label{tab2}}
\end{center}
\end{table}

\subsection{Split bundles and stability}\label{split_stab}
We would like to construct vector bundles $V$ with the required structure group $H=S(U(n_1)\times\dots\times U(n_f))$. Starting with vector bundles $U_a$ on the Calabi-Yau three-fold $X$, each with structure group  $U(n_a)$, we set
\begin{equation}
 V=\bigoplus_{a=1}^f U_a\; . \label{Vdef}
\end{equation}
In order to ensure that the structure group is {\em special} unitary we also impose the vanishing of the first Chern class~\footnote{\label{footnote4}If there are additional conditions between the first Chern classes of the $U_a$ the structure group might reduce further and become a proper sub-group of one of the structure groups given in Table~\ref{tab1}. In this case, the non-Abelian part of the low-energy gauge group might be larger than $SU(5)$. We will not consider this case explicitly in our general set-up and avoid models of this type in our discussion of examples later on.}
\begin{equation}
 c_1(V)=\sum_{a=1}^{f}c_1(U_a)\stackrel{!}{=}0\; . 
\end{equation}
Relative to a basis $\{\omega_i\}$ of harmonic two-forms on $X$, where $i=1,\ldots ,h^{1,1}(X)$, we expand the first Chern classes as $c_1(U_a)=c_1^i(U_a)\omega_i$ and, in order to make contact with the four-dimensional discussion in the previous section, introduce the vectors ${\bf k}_a$ by setting
\begin{equation}
 k_a^i=c_1^i(U_a) \label{kdef}\; .
\end{equation}
Now we need to discuss the conditions on such bundles $V$ which follow from the requirement of preserving four-dimensional $N=1$ supersymmetry. As discussed above, in order for this bundle to be supersymmetric, it needs to be poly-stable with zero slope.
 
To understand these conditions, we must define the notions of slope, stability and poly-stability which we introduce in turn. The slope $\mu({\cal G})$ of a coherent sheaf, ${\cal G}$, on the Calabi-Yau three-fold $X$ is defined by
\begin{equation}
 \mu({\cal G})=\frac{1}{{\rm rk}({\cal G})}\int_X c_1({\cal G})\wedge J\wedge J=\frac{1}{{\rm rk}({\cal G})}d_{ijk}c_1^i({\cal G})t^jt^k\; , \label{mudef}
\end{equation}
where $J$ is the Kahler form of $X$.  For the second equality we have expanded $J=t^i\omega_i$ and $c_1({\cal G})=c_1^i({\cal G})\omega_i$ and introduced the triple intersection numbers $d_{ijk}=\int_X\omega_i\wedge\omega_j\wedge\omega_k$ of $X$. 
A holomorphic vector bundle $V$ is now called (slope-) {\it stable} if 
\beq
\mu({\cal G})<\mu(V)~~\text{for all coherent sub-sheaves}~{\cal G}\subset V~~\text{with}~0<{\rm rk}({\cal G})<{\rm rk}(V)\; .
\eeq 
Note that due to the restriction on the rank in this definition line bundles are always stable. Further, $V$ is called poly-stable if
\beq
V= \bigoplus_a V_a~~\text{such that}~V_a~\text{stable and}~~\mu(V_a)=\mu(V)~ \forall a \label{Vsum}
\eeq
Hence, a poly-stable bundle consists of a direct sum of stable bundles, each with the same slope. Since supersymmetry also requires that $\mu(V)=0$ (which is automatic in our case since we consider bundles $V$ with $c_1(V)=0$) this means that the slope of all constituent bundles $V_a$ must vanish.
A poly-stable bundle $V$ with zero slope has no global sections since the trivial line bundle ${\cal O}_X$ has slope zero and is, hence, a potentially de-stabilising sub-bundle. Therefore, ${\cal O}_X$ cannot inject into $V$ and we must have $H^0(X,V)\cong{\rm Hom}({\cal O}_X,V)=0$. If $V$ is poly-stable, its dual $V^*$, is also poly-stable with zero slope, so that on a Calabi-Yau manifold $H^3(X,V)\cong H^0(X,V^*)=0$. In conclusion, poly-stable bundles with slope zero on a Calabi-Yau manifold have vanishing zeroth and third cohomology\footnote{\label{footnote5}If a line bundle appears in the sum~\eqref{Vsum} the above argument breaks down. However, in this case, one can still conclude that a poly-stable, zero slope bundle $V$ satisfies $H^0(X,V)=H^3(X,V)=0$ by invoking the vanishing Theorem (1.24) in Ref.~\cite{kob}. It states that a line bundle $L$ has no global sections if $\mu(L)<0$ (and, on a Calabi-Yau manifold, it has vanishing third cohomology if $\mu(L)>0$) somewhere in the Kahler cone of $X$. Since we require that $\mu(L)=0$, all line bundles except the trivial one have points in the Kahler cone where $\mu(L)>0$ and $\mu(L)<0$, so that the theorem applies and $H^0(X,L)=H^3(X,L)=0$. The one exception is the trivial bundle $\cO_X$ which has vanishing slope everywhere in the Kahler cone and satisfied $H^0(X,\cO_X)=H^3(X,\cO_X)=1$. However, we are not interested in cases for which $\cO_X$ appears in the direct sum in \eref{Vdef} since this leads to the case of enhanced symmetry described in footnote \ref{footnote4}. Hence, for our considerations, all line bundles $L$ will indeed have vanishing zeroth and third cohomology.}.

Let us now apply these general statements to the bundle $V$ in \eqref{Vdef}. For $V$ to be supersymmetric it needs to be poly-stable with zero slope which is equivalent to saying that all $U_a$ must be stable with zero slope, $\mu(U_a)=0$. Stability is automatic if $U_a$ is a line bundle but has to be checked explicitly for higher-rank bundles (this can be carried out explicitly following the procedure outlined in Ref.~\cite{Anderson:2008ex,Anderson:2009nt}). 
For the full bundle $V$ to be poly-stable, the stable loci for the various $U_a$ must have a non-trivial intersection in the Kahler cone. On this intersection $V$ is poly-stable. The second condition for supersymmetry, the vanishing of the slope for each $U_a$, can from Eq.~\eqref{mudef} be expressed as
\begin{equation}
 \mu(U_a)\sim k^i_a\kappa_i=0~~\forall a\; , \label{slopecond}
\end{equation}  
where $\kappa_i=d_{ijk}t^jt^k$ are ``dual" Kahler moduli space coordinates. Hence, the vanishing slope conditions lead to additional constraints on the Kahler moduli space of $X$ which have to be combined with the ones following from stability. As can be seen from Eq.~\eqref{Dterm}, in the four-dimensional effective theory all these constraints are enforced via D-terms associated to the anomalous $U(1)$ symmetries in $\anU$ and additional anomalous $U(1)$ symmetries which may appear at particular loci in Kahler moduli space when one or more of the bundles $U_a$ split up further~\cite{Anderson:2010tc}. The bundle is supersymmetric only in the part of the Kahler moduli space where all of these conditions are satisfied simultaneously. As is clear from Eq.~\eqref{slopecond}, in order for a common solution to the zero slope conditions to exist it is necessary that
\begin{equation}
 (\mbox{number of lin.~independent } {\bf k}_a)<h^{1,1}(X)\; .\label{numconsgen}
\end{equation}
In the context of the four-dimensional discussion we have seen the same condition~\eqref{numcons}, for the case of purely Abelian splittings, appear from the D-term equations.

For Calabi-Yau three-folds with a small Hodge number $h^{1,1}(X)$, the slope zero conditions~\eqref{slopecond} are an important model building constraint on the bundle, $V$. For example, for $h^{1,1}(X)=1$ there are no solutions at all, while for $h^{1,1}(X)=2$ all first Chern classes ${\bf k}_a$ must be multiples of each other. 

As we have seen, stability and vanishing slope of each $U_a$ implies that $H^0(X,U_a)=H^3(X,U_a)=0$. This means the chiral asymmetries associated to the bundles $U_a$ (that is the chiral asymmetry of the ${\bf 10}$ and $\bar{\bf 10}$ multiplets with charges $\pm{\bf e}_a$) can be computed from the index, so that ${\rm ind}(U_a)=-h^1(X,U_a)+h^2(X,U_a)$.

An analogous argument holds for $\wedge^2 V$ and its index. To see this, note that if $V$ is a poly-stable bundle with slope zero then it follows that $\wedge^2 V$ (and $V\otimes V^*$) are also poly-stable with slope zero \cite{friedman}. As a result, each indecomposable term, $ \wedge^2 U_a, U_a \otimes U_b$ in $\wedge^2 V$ is a properly stable bundle with slope zero. Following the same line of argument once again, such a term either has vanishing zeroth and third cohomology, or consists of a trivial bundle in which case its zeroth and third cohomologies are equal. Either way we have that ${\rm ind}(\wedge^2 U_a)=-h^1(X,\wedge^2 U_a)+h^2(X,\wedge^2U_a)$ (and similarly for $U_a\otimes U_b$), so that the index counts the chiral asymmetry of $\bar{\bf 5}$ and ${\bf 5}$ multiplets with charges $\pm({\bf e}_a+{\bf e}_b)$.

\subsection{Spectrum of GUT theory}

As discussed above, the four-dimensional gauge group is $SU(5)\times \anU$, where $\anU\simeq U(1)^{f-1}$ consists of $U(1)$ factors which are normally Green-Schwarz anomalous. From Table~\ref{tab2}, the multiplets $r_{\bf q}$ under this gauge group are given by
\begin{equation}
  {\bf 1}_{{\bf e}_a-{\bf e}_b},\bar{\bf 5}_{{\bf e}_a+{\bf e}_b},{\bf 10}_{{\bf e}_a},{\bf 5}_{-{\bf e}_a-{\bf e}_b}, \label{mult}
   \overline{\bf 10}_{-{\bf e}_a}
\end{equation} 
where we recall that the sub-script indicates the $\anU$ charge, an integer vector subject to the identification~\eqref{qid1}. The corresponding internal $H$ representation, $R_{\bf q}$, for each of these multiplets is listed in the first column of Table~\ref{tab2}. Given that the bundles $U_a$ are associated to the $H$-representations $({\cal F}_a)_{{\bf e}_a}$, the associated bundles for each of the multiplets $r_{\bf q}$ in~\eqref{mult} are easily worked out by first identifying the corresponding $H$ representation $R_{\bf q}$ and then taking appropriate tensor products of $({\cal F}_a)_{{\bf e}_a}$. The result for the associated bundles $U_{R_{\bf q}}$ is listed in the third column of Table~\ref{tab2}. The number, $n(r_{\bf q})$ of multiplets with $SU(5)\times \anU$ representation $r_{\bf q}$ is then given by the first cohomology, $H^1(X,U_{R_{\bf q}})$, of these associated bundle. Serre duality on a Calabi-Yau manifold implies that $H^1(X,U)\cong H^2(X,U^*)$ for any bundle $U$. Hence, the chiral asymmetries
$N(r_{\bf q})=n(r_{\bf q})-n(\bar{r}_{-{\bf q}})$ can be expressed in terms in terms of the topological index of each respective bundle as
\begin{eqnarray}
 N({\bf 1}_{{\bf e}_a-{\bf e}_b})&=&n({\bf 1}_{{\bf e}_a-{\bf e}_b})-n({\bf 1}_{{\bf e}_b-{\bf e}_a})=-{\rm ind}(U_a\otimes U_b^*)\label{ii}\\
 &=&-{\rm rk}(U_b){\rm ind}(U_a)+{\rm rk}(U_a){\rm ind}(U_b)-{\rm ch}_1(U_a){\rm ch}_2(U_b)+{\rm ch}_2(U_a){\rm ch}_1(U_b)\nn\\
 N(\bar{\bf 5}_{2{\bf e}_a})&=&n(\bar{\bf 5}_{2{\bf e}_a})-n({\bf 5}_{-2{\bf e}_a})=-{\rm ind}(\wedge^2 U_a)\\
 &=&(4-{\rm rk}(U_a)){\rm ind}(U_a)-{\rm ch}_1(U_a)\left({\rm ch}_2(U_a)+\frac{1}{4}c_2(TX)\right)\\
 N(\bar{\bf 5}_{{\bf e}_a+{\bf e}_b})&=&n(\bar{\bf 5}_{{\bf e}_a+{\bf e}_b})+n({\bf 5}_{-{\bf e}_a-{\bf e}_b})=-{\rm ind}(U_a\otimes U_b)\\
 &=&-({\rm rk}(U_b){\rm ind}(U_a)+{\rm rk}(U_a){\rm ind}(U_b)+{\rm ch}_1(U_a){\rm ch}_2(U_b)+{\rm ch}_2(U_a){\rm ch}_1(U_b))\nn\\
 N({\bf 10}_{{\bf e}_a})&=&n({\bf 10}_{{\bf e}_a})-n(\overline{\bf 10}_{-{\bf e}_a})=-{\rm ind}(U_a)\; .\label{if}
\end{eqnarray} 
The total chiral asymmetry of $\bar{\bf 5}$ and ${\bf 10}$ multiplets, summed over all $\anU$ charges, is given by the usual formula
\begin{equation}
 N({\bf 10}) =-{\rm ind}(V)=-{\rm ind}(\wedge^2 V)=N(\bar{\bf 5}) \label{indequal}
\end{equation}
for poly-stable rank five bundles with $c_1(V)=0$.
\subsection{Discrete symmetries, equivariance and downstairs spectrum}
\label{sec:symm}
As in generic heterotic compactifications, in order to break the visible GUT symmetry to $G_{\rm SM}=SU(3)\times SU(2) \times U_Y(1)$, we must introduce Wilson lines, on a non-simply connected space. To this end, we will quotient the Calabi-Yau three-fold and bundle by a freely-acting discrete symmetry $\Gamma$. A bundle $U$ over $X$ descends to a bundle $\hat{U}$ on the quotient $\hat{X}=X/\Gamma$ if and only if it is equivariant, that is, if the symmetry $\Gamma$ can be ``lifted" to the bundle. The mathematical definitions and the details on how to construct equivariant structures are described in Appendix~\ref{appA}. Here we will merely need a few facts about such bundles. First, the indices of an equivariant bundle $U$ and its downstairs counterpart $\hat{U}$ are related by
\begin{equation}
{\ind}(\hat{U})={\rm ind}(U)/|\Gamma|\; . \label{inddiv}
\end{equation}
Also the cohomologies, $H^q(X,U)$, of an equivariant bundle $U$ form representations under the group $\Gamma$. It is therefore useful to define the ``graded" cohomologies $H^q(X,U,R)$ which are the subspaces of $H^q(X,U)$ which transform under the $\Gamma$ representations $R$. Further, $h^q(X,U,R)={\rm dim}( H^q(X,U,R))$ is the corresponding graded dimension and the graded index is defined by
\begin{equation}
 {\rm ind}(U,R)=\sum_q(-1)^qh^q(X,U,R)\; .\label{indcond}
\end{equation}

As explained earlier, we focus on models where each constituent bundle $U_a$ has an equivariant structure on its own. The obvious ambiguity in choosing such equivariant structures consists of an overall phase for each $U_a$ which can be encoded in $\Gamma$-representations or characters, denoted by $\chi_a^*$. We should now discuss how to obtain the downstairs spectrum. Eq.~\eqref{inddiv} applied to each $U_a$ means that the downstairs spectrum consists of a certain number of complete $\bar{\bf 5}$ and ${\bf 10}$ representations plus vector-like multiplets. In particular, each downstairs $\bar{\bf 5}$ ({\bf 10}) multiplet descends as a whole from a specific sector of $\anU$ charge. This means that the constituents of each such multiplet carry the same charge under the anomalous $U(1)$ symmetries in $\anU$. 

In order to find the vector-like spectrum downstairs one has to study the equivariant structure in more detail. To do so, we must choose a Wilson line ${\cal W}$ (that is, a flat rank one bundle on $\hat{X}$, embedded into $U_Y(1)$ in order to break the GUT symmetry to the standard model) which induces a representation of the discrete group $\Gamma$ contained in $U_{Y}(1)$. Each standard model multiplet, $\psi$, in Table~\ref{tab2} is, therefore, associated to an Wilson line bundle ${\cal W}_\psi$ and carries a $\Gamma$ representation $R_\psi$ which is determined by its weak hypercharge. If we focus on a particular such  multiplet, $\psi$, contained within the GUT multiplet associated to $U$ then we have the following relation between cohomologies.
\begin{equation}
 H^1(\hat{X},\hat{U}\oplus {\cal W}_\psi)\cong (H^1(X,U)\otimes R_\psi)_{\rm inv}
\end{equation} 
That is, the downstairs spectrum in the presence of the Wilson line can be computed from the upstairs cohomology by tensoring with the various $\Gamma$ representations of the standard model multiplets and extracting the $\Gamma$ invariant part from this tensor product.

Let us now explain this procedure in more detail, focusing on our models with split structure group. Here we will only discuss Abelian discrete symmetries
\begin{equation} 
 \Gamma=\bigotimes_r\mathbb{Z}_{m_r}\; ,
\end{equation}
typically with either a single or with two $\mathbb{Z}_m$ factors. We define $\alpha_r=\exp (2\pi i/m_r)$ and the $\mathbb{Z}_{m_r}$ representations $R_{p_r}^{(m_r)}(g)=\alpha_r^{p_r g}$, where $g\in\{0,\ldots ,m_r-1\}$. Then, an embedding of $\mathbb{Z}_{m_i}$ into $U_Y(1)\subset SU(5)$ can be written as
 \begin{equation}
 g\rightarrow {\rm diag}(\alpha_r^{p_rg},\alpha_r^{p_rg},\alpha_r^{p_rg},\alpha_r^{\tilde{p}_rg},\alpha_r^{\tilde{p}_rg})\; ,
\end{equation} 
where $p_r$, $\tilde{p}_r$ are integers satisfying $3p_r+2\tilde{p}_r=0\mbox{ mod } m_r$. Further, to actually break $SU(5)$ to $G_{\rm SM}$ we need that $p_r\neq \tilde{p}_r$ for at least one $r$. So in summary, the viable $SU(5)$ breaking Wilson lines can be obtained by solving
\begin{equation}
 3p_r+2\tilde{p}_r=0\mbox{ mod } m_r\mbox{ for all }r\; ,\quad p_i\neq \tilde{p}_r\mbox{ for at least one }r\; , \label{Wilsoncons}
 \end{equation}
 where $p_r, \tilde{p}_r\in \{0,\ldots ,m_r-1\}$. In particular, this means that a single $\mathbb{Z}_m$ can break $SU(5) $ to the standard model provided that $n\neq 5$. If $\Gamma$ is a direct product of $\mathbb{Z}_m$ factors is it sufficient that one of the factors is different from $\mathbb{Z}_5$.
For a solution of the above equations, we define the two $\Gamma$ representations $W=\bigotimes_rR_{p_r}^{(m_r)}$ and $\tilde{W}=\bigotimes_rR_{\tilde{p}_r}^{(m_r)}$. The relevant $SU(5)$ representations decompose under $\Gamma\times G_{\rm SM}$ as
\begin{eqnarray}
 {\bf 5}&\rightarrow&(\tilde{W},{\bf 1},{\bf 2})_3\oplus(W,{\bf 3},{\bf 1})_{-2}\\
 \bar{\bf 5}&\rightarrow& (\tilde{W}^*,{\bf 1},{\bf 2})_{-3}\oplus(W^*,\bar{\bf 3},{\bf 1})_2\\
 {\bf 10}&\rightarrow&(\tilde{W}\otimes\tilde{W},{\bf 1},{\bf 1})_6\oplus(W\otimes W,\bar{\bf 3},{\bf 1})_{-4})\oplus (W\otimes\tilde{W},{\bf 3},{\bf 2})_1\\
 \bar{\bf 10}&\rightarrow&(\tilde{W}^*\otimes\tilde{W}^*,{\bf 1},{\bf 1})_{-6}\oplus(W^*\otimes W^*,{\bf 3},{\bf 1})_{4})\oplus (W^*\otimes\tilde{W}^*,\bar{\bf 3},{\bf 2})_{-1}\; .
\end{eqnarray} 
Then, the downstairs cohomology can be expressed in terms of graded cohomologies with respect to the $\Gamma$ representations $W$, $\tilde{W}$, which describe the Wilson line and the characters $\chi_a$, which encode the freedom in choosing the equivariant structure. The result is summarized in Table~\ref{tab:gc}.
\begin{table}
\begin{center}
\begin{tabular}{|l|l|l|l|}\hline
 $SU(5)$ repr.&$G_{\rm SM}$ repr.&name&cohomology\\\hline\hline
 ${\bf 10}_{{\bf e}_a}$&$({\bf 3},{\bf 2})_1$&$Q_a$&$h^1(X,U_a,\chi_a\otimes W^*\otimes\tilde{W}^*)$\\
 &$(\bar{\bf 3},{\bf 1})_{-4}$&$u_a$&$h^1(X,U_a,\chi_a\otimes W^*\otimes W^*)$\\
 &$({\bf 1},{\bf 1})_6$&$e_a$&$h^1(X,U_a,\chi_a\otimes\tilde{W}^*\otimes\tilde{W}^*)$\\\hline
 $\bar{\bf 5}_{{\bf e}_a+{\bf e}_b}$&$(\bar{\bf 3},{\bf 1})_2$&$d_{a,b},\, T_{a,b}$&$h^1(U_a\otimes U_b,\chi_a\otimes \chi_b\otimes W)$\\
 &$({\bf 1},{\bf 2})_{-3}$&$L_{a,b},\, H_{a,b}$&$h^1(U_a\otimes U_b,\chi_a\otimes\chi_b\otimes\tilde{W})$\\\hline
 $\bar{\bf 5}_{2{\bf e}_a}$&$(\bar{\bf 3},{\bf 1})_2$&$d_a,\,T_a$&$h^1(\wedge^2U_a,\chi_a\otimes \chi_a\otimes W)$\\
 &$({\bf 1},{\bf 2})_{-3}$&$L_{a},\, H_{a}$&$h^1(\wedge^2 U_a,\chi_a\otimes\chi_a\otimes\tilde{W})$\\\hline
 ${\bf 5}_{-{\bf e}_a-{\bf e}_b}$&$({\bf 3},{\bf 1})_{-2}$&$\bar{T}_{a,b}$&$h^2(U_a\otimes U_b,\chi_a\otimes \chi_b\otimes W)$\\
 &$({\bf 1},{\bf 2})_3$&$\bar{H}_{a,b}$&$h^2(U_a\otimes U_b,\chi_a\otimes\chi_b\otimes\tilde{W})$\\\hline
 
${\bf 5}_{-2{\bf e}_a}$&$({\bf 3},{\bf 1})_{-2}$&$\bar{T}_{a}$&$h^2(\wedge^2U_a,\chi_a\otimes \chi_a\otimes W)$\\
 &$({\bf 1},{\bf 2})_3$&$\bar{H}_{a}$&$h^2(\wedge^2U_a,\chi_a\otimes\chi_a\otimes\tilde{W})$\\\hline
 
 ${\bf 1}_{{\bf e}_a-{\bf e}_b}$&$({\bf 1},{\bf 1})_0$&$S_{a,b}$&$h^1(U_a\otimes U_b^*,\chi_a\otimes \chi_b^*)$\\\hline
\end{tabular}
\parbox{5in}{\caption{\it\small Cohomologies which compute the downstairs spectrum. The Wilson line is characterized by the two representations $W$ and $\tilde{W}$ of the freely-acting Abelian symmetry $\Gamma$ and $\chi_a^*$ are the characters of $U_a$. The number of mirror particles is obtained by the second cohomology of the same bundle and the same representation.}\label{tab:gc}}
\end{center}
\end{table}

For the ``physics" models we consider in this paper not all of the graded cohomologies in Table~\ref{tab:gc} need to be computed explicitly. The upstairs spectrum of such models consists of $3|\Gamma|$ ${\bf 10}$ multiplets and has no $\bar{\bf 10}$ multiplets. By virtue of Eq.~\eqref{inddiv} this will guarantee precisely three ${\bf 10}$ multiplets downstairs so there is no need to check graded cohomologies in this sector. Further, from~Eq.~\eqref{indequal} we know that the chiral asymmetry of $\bar{\bf 5}$ and ${\bf 5}$ multiplets is also $3|\Gamma|$, so that we are guaranteed three chiral $\bar{\bf 5}$ multiplets downstairs. In addition, we have to check that all Higgs triplets can be projected out and at least one pair of Higgs doublets remains in the spectrum. This can be done by computing the number of $\bar{T}_{a,b}$ and $\bar{H}_{a,b}$ multiplets from the associated cohomologies in Table~\eqref{tab:gc}. The details of how to compute graded cohomologies for line bundles on the particular Calabi-Yau manifolds used in our constructions are explained in Appendix~\ref{appB}.
\subsection{Anomalies}
A characteristic feature of our split models is the presence of the Green-Schwarz anomalous $U(1)$ symmetries in $\anU$. For the case of line bundle sums and from a four-dimensional perspective this has already been discussed in Section~\ref{sec:gs4d}. Here, we will provide a general discussion, valid for all splitting types and from a geometric viewpoint.

We begin by computing the anomaly coefficients in the GUT model, focusing on the mixed $\anU\,SU(5)^2$ contribution.
With the group theoretical indices $c(\bar{\bf 5})=1$ and $c({\bf 10})=3$ for the relevant $SU(5)$ representations the $\anU\,SU(5)^2$ triangle anomaly is proportional to
\begin{equation}
 {\bf A}=\sum_{a=1}^f\left(2N(\bar{\bf 5}_{2{\bf e}_a})+3N({\bf 10}_{{\bf e}_a})\right) {\bf e}_a+\sum_{a<b}N(\bar{\bf 5}_{{\bf e}_a+{\bf e}_b})({\bf e}_a+{\bf e}_b)\; . \label{anomaly1}
\end{equation} 
Here, we recall that $N(r_{\bf q})=n(r_{\bf q})-n(\bar{r}_{-{\bf q}})$ is the chiral asymmetry of a certain representation $r_{\bf q}$. The above anomaly coefficient is an $f$-dimensional vector which is defined only up to the identification~\eqref{qid1} of $\anU$ charges, so we should explain how to extract unambiguous anomaly coefficients from this result. Any particular $U(1)$ symmetry within $\anU$ can be represented by a vector ${\bf Q}=(Q_1,\ldots, Q_f)$ satisfying ${\bf n}\cdot {\bf Q}=0$. The anomaly coefficient for this $U(1)$ symmetry is then given by ${\bf Q}\cdot{\bf A}$ and is, hence, independent of the identification~\eqref{qid1}. 

We can use the results~\eqref{ii}--\eqref{if} for the chiral asymmetries to rewrite the above expression for ${\bf A}$ in terms of topological data. This leads to
\begin{equation}
 (A_a-\beta_i k_a^i)_{a=1,\ldots ,f}\in\mathbb{Z}{\bf n}\; ,\qquad \beta_i=c_{2i}(V)-\frac{1}{2}c_{2i}(TX)\; , \label{anomaly2}
\end{equation} 
where $\beta$ is the coefficient which appears in the one-loop correction to the four-dimensional gauge-kinetic function~\eqref{f}.
We have phrased the above results for the upstairs Calabi-Yau $X$ and the upstairs bundle $V$. However, analogous  equations, with the chiral asymmetries in~\eqref{anomaly1} interpreted as the downstairs chiral asymmetries and the replacements $U_a\rightarrow \hat{U}_a$, $V\rightarrow\hat{V}$ and $X\rightarrow \hat{X}$ in \eqref{anomaly2} hold on the quotient Calabi-Yau $\hat{X}$, provided all bundles $U_a$ are equivariant individually, as we are assuming here. The subsequent inclusion of a Wilson line does, of course, not affect the chiral asymmetries and, hence, leaves the anomaly coefficients unchanged.

It is important to also consider the masses of the $f-1$ vector fields associated to the potentially anomalous $U(1)$ symmetries in $\anU$. They are given by Eq.~\eqref{U1masses}, but with the more general definition~\eqref{kdef} of the first Chern classes. The mass matrix~\eqref{U1masses} has size $f$ while we only have $f-1$ vector fields. As usual, this mismatch is corrected by imposing the constraint ${\bf n}\cdot{\bf v}=0$ on $f$ dimensional vectors ${\bf v}=(v^a)$ representing gauge bosons in this basis. Hence, following the discussion in Section~\ref{sec:U1masses}, massless vector $U(1)$ bosons are characterized by the equation
\begin{equation}
 \sum_{a=1}^{f}{\bf k}_av^a=0\mbox{ for } {\bf n}\cdot{\bf v}=0\; . \label{U1massless}
\end{equation} 
Comparison with Eq.~\eqref{anomaly2} shows that a $U(1)$ symmetry with a massless vector boson is necessarily anomaly-free. Conversely, while a $U(1)$ symmetry with a massive vector boson can generically be expected to be anomalous, it may be anomaly-free in special cases. This happens if $k_a^iv^a\beta_i=0$ while $k_a^iv^a\neq 0$. This means the number of massless $U(1)$ vector bosons is given by
\begin{equation}
 (\mbox{number of massless }U(1)\mbox{ vector bosons})=f-1-{\rm rank}(k_a^i)\; ,
\end{equation} 
and combining this result with~\eqref{numconsgen} we learn that
\begin{equation}
(\mbox{number of massless }U(1)\mbox{ vector bosons})\geq f-h^{1,1}(X)\; .
\end{equation}
This means that for Calabi-Yau manifolds with small $h^{1,1}(X)<f$ we necessarily have at least one massless $U(1)$ vector boson. However it is frequently still possible to spontaneously break these symmetries by giving a vacuum expectation value to appropriate singlet fields. The phenomenological implications of such choices were explored in Section \ref{sec:database}. In the next section, we will explore the geometric meaning of such singlet VEVs.

\subsection{Connecting different splitting types}
\label{sec:dtermsgen}

Split bundles of the type considered in this paper typically arise at a special locus in Kahler and bundle moduli space and, provided certain fields exist, one can move away from this special locus to obtain a less-split bundle or even recover a full $SU(5)$ structure group. This process corresponds to moving upwards in the list of structure groups given in Table~\ref{tab1}. Conversely, setting certain moduli to zero one can go back to the special locus and obtain a finer splitting of the structure group. Such movement in moduli space has been studied in detail in recent literature \cite{Sharpe:1998zu,Kuriyama:2008pv,Anderson:2009sw,Anderson:2009nt,Anderson:2010tc,Anderson:2010ty} and here we will simply review some of the relevant facts as they apply to the models under investigation.

The moduli of the split bundle, $V$, in Eq.~\eqref{Vdef} are described by
\begin{equation}
 H^1(X,V\otimes V^*)=\bigoplus_{a,b}H^1(X,U_a\otimes U_b^*)\; . \label{Vmod}
\end{equation} 
The $a=b$ terms on the right-hand side clearly correspond to moduli of the bundle $U_a$. It is suggestive (and in fact correct) to relate the terms for $a\neq b$ to deformations away from the split bundles $U_a\oplus U_b$. To see this in more detail, let us focus on two constituent bundles $U_a$, $U_b$ with structure groups $U(n_a)$, $U(n_b)$. Possible bundles $U$ which correspond to moving away from the split bundle $U_a\oplus U_b$ can be described by the following two extension sequences
\begin{align}
& 0\rightarrow U_a\rightarrow U\rightarrow U_b\rightarrow 0\label{ext1}\; . \\\label{ext2}
& 0 \rightarrow U_b \rightarrow \tilde{U} \to U_a \to 0
\end{align} 
Any infinitesimal smooth, slope-stable deformation of $U_a \oplus U_b$ can be described by a combination of the bundles $U$ and $\tilde{U}$ above \cite{Li:2004hx}. For example, the space of inequivalent bundles $U$ (respectively $\tilde{U}$) described by this sequence is given by ${\rm Ext}^1(U_b,U_a)\cong H^1(X,U_a\otimes U_b^*)$ (respectively ${\rm Ext}^1(U_a, U_b) \cong H^1(X, U_b \otimes U_a^*)$), so precisely the terms for $a\neq b$ in the sum~\eqref{Vmod}. Comparing with Table~\ref{tab2} we see that these cohomology groups are associated to the $SU(5)$ singlet fields $S_{ab}$ which carry charge ${\bf e}_a-{\bf e}_b$ under the anomalous $U(1)$ symmetries in $\anU$. Hence, once we have computed the complete spectrum of a given split bundle $V$ we can immediately decide which deformations are possible. 

If $H^1(X,U_a\otimes U_b^*)= H^1(X,U_b\otimes U_a^*)=0$ then no relevant singlet fields exist and the split bundle $U=U_a\oplus U_b$ is the only bundle possible in the extension sequences \eqref{ext1}, \eref{ext2}. On the other hand, if at least one of $H^1(X,U_a\otimes U_b^*)$ and $H^1(X,U_b\otimes U_a^*)$ is non-zero, non-trivial extensions $U$ whose structure group is ``larger" than $U(n_a)\times U(n_b)$ and is contained in $U(n_a+n_b)$ exist. 

In the four-dimensional effective theory these infinitesimal deformations are described by the D-terms associated to the anomalous $U(1)$ symmetries. For the case of line bundle sums this has been discussed in Section~\ref{sec:Dterms}. For general splittings and after setting all non-singlet fields to zero, the schematic structure of these D-terms is
\begin{equation}
 D_a=\frac{3 \mu(U_a)}{\kappa}+\frac{1}{n_a}\sum_{\stackrel{b}{b\neq a}}\left(|S_{ab}|^2-|S_{ba}|^2\right)\; , \label{Dterms2}
\end{equation}
where $S_{ab}$ are bundle moduli singlets, ${\kappa=d_{ijk}t^it^jt^k}$ is the Calabi-Yau volume and the slope, $\mu$, has been defined in \eqref{mudef}~\footnote{For simplicity of notation we have dropped possible Kahler metrics in front of the singlet matter field terms. Also, note that, in order to keep our notation more ``covariant", we have written down $f$ D-terms for only $f-1$ anomalous $U(1)$ symmetries. This does not lead to an additional constraint since $\sum_a n_a D_a=0$, as a consequence of $c_1(V)=\sum_ac_1(U_a)=0$. The actual four-dimensional D-terms are $f-1$ linearly independent combinations of the $D_a$.}. Given these expressions for the D-terms, we can now interpret the geometric discussion above from a four-dimensional point of view. 

For a split bundle $U=U_a\oplus U_b$ all the relevant singlet fields (if they even exist) should be zero in vacuum, that is, $\langle S_{ab} \rangle =\langle S_{ba} \rangle=0$. In this case, in order to satisfy the D-term equations, we should require that $\mu(U_a)=\mu(U_b)=0$. This corresponds precisely to the zero-slope poly-stability conditions \eqref{slopecond} on the split bundle which constrain the Kahler moduli. If no fields $S_{ab}$ or $S_{ba}$ exist this is the only way to satisfy the D-term equations. In this case, some of the Kahler moduli are stabilised and the bundle is necessarily split. On the other hand, if fields $S_{ab}$ or $S_{ba}$ do exist they may be given a vacuum expectation value, provided this is consistent with D- and F-flatness of the four-dimensional  theory. This corresponds to a non-trivial extension $U$ (respectively, $\tilde{U}$) of $U_a\oplus U_b$, with structure group $U(n_a+n_b)$. Solving the D-term equations then typically requires that $\mu(U_a)\neq 0$ and $\mu(U_b)\neq 0$, that is, at the same time as moving in the bundle moduli space one is forced to move in the Kahler moduli space away from the special zero-slope locus. Due to the non-trivial singlet VEVs some or all of the anomalous $U(1)$ symmetries are broken spontaneously. While this merely generates an additional breaking for the $U(1)$ symmetries which are anomalous this can give masses to the $U(1)$ symmetries which have remained massless at the split locus. Also, note that the direction in which one can move away from the slope-zero locus is dictated by which singlet fields are present. If there are fields $S_{ab}$ with $a<b$ only, then one can only leave the locus in directions characterised by $\mu(U_a)<0$ and $\mu(U_b)>0$. For fields $S_{ba}$ only we have the opposite inequalities and, finally, if both types of fields are present there need not be a restriction in Kahler moduli space. 

One must also be careful in this context to consider the constraints arising from bundle holomorphy. While these conditions are satisfied automatically by ensuring that the component bundles $U_a$ are holomorphic, it is not automatic that all choices of singlet VEVs lead to holomorphic bundles. These obstructions to the deformations appear as non-trivial F-terms in the four-dimensional theory, as discussed in Sections \ref{tea} and \ref{sec:datafterms}. Such F-term obstructions must be explored on a case-by-case basis (see \cite{Berglund:1995yu,Anderson:2010tc,Anderson:2010ty,Anderson:2010mh,Anderson:2011ty} for further discussions).

In summary, depending on the choice of bundle $V$, various or possibly all of the different splitting types in Table~\ref{tab1} may be connected by deformations and may, hence, be part of the same moduli space. Whether bundles can be connected in such a way, that is, whether one can move away from a given zero-slope locus in Kahler moduli so that the $U(1)$ symmetries are spontaneously broken, is of obvious relevance for the phenomenology of our models. It is important that we can answer these questions, not only by exploring the effect of singlet VEVs in the four-dimensional effective theory, but also explicitly by a direct calculation of the cohomology groups $H^1(X,U_a\otimes U_b^*)$. 

With this brief overview of deformation theory in place, we have completed our exploration of the consistency conditions on split gauge bundles in heterotic compactifications. We briefly summarize these conditions below, before turning to phenomenologically interesting examples.

\subsection{The conditions imposed on the geometry}
\subsubsection{Consistency conditions on the bundles}
\label{sec:consistbund}

Here we will concisely summarize the conditions for a consistent heterotic compactification discussed in the previous section. We assume that we have given a Calabi-Yau manifold $X$ with a freely-acting discrete symmetry $\Gamma$. In order to be able to break $SU(5)$ to the standard model group, for $\Gamma=\mathbb{Z}_n$ we should have $n\neq 5$ and for $\Gamma$ given by a product of $\mathbb{Z}_n$ factors at least one factor should be different from $\mathbb{Z}_5$. As discussed earlier, such a Wilson line can be characterized by two $\Gamma$ representations $W$ and $\tilde{W}$, subject to the constraints~\eqref{Wilsoncons}. Further, on the Calabi-Yau manifold $X$, we have a vector bundle $V=\oplus_{a=1}^fU_a$ with $c_1(V)=0$, where $U_a$ are bundles with structure group $U(n_a)$ and first Chern class $k_a^i=c_1^i(U_a)$. Then, the full structure group is given by $H=S(U(n_1)\times\dots\times U(n_f))$ and will frequently be referred to by the split vector ${\bf n}=(n_1,\ldots ,n_f)$. On this data we impose the following conditions.
\begin{itemize}
\item[(C1)] There should be an anomaly-free completion of the model, that is we should be able to satisfy Eq.~\eqref{ancond}. In practice, we will guarantee this by requiring that
\begin{equation}
 c_2(TX)-c_2(V)\stackrel{!}{\in}\mbox{Mori cone of }X\; ,
\end{equation} 
so that a complete model can be obtained by adding a suitable number of five-branes and choosing the hidden bundle to be trivial.
\item[(C2)] The bundle $V$ must be holomorphic and poly-stable with zero slope, somewhere in the Kahler cone of $X$. As discussed, the second of these constraints means that every constituent bundle $U_a$ is stable with vanishing slope and that the resulting constraints on the Kahler moduli have a common solution. In particular, all of the zero slope conditions
\begin{equation}
 \mu(U_a)\sim d_{ijk}k^i_at^jt^k=k_a^i\kappa_i\stackrel{!}{=}0
\end{equation}
should have a common solutions. This necessarily means that the number of linearly independent Chern classes ${\bf k}_a$ is less than $h^{1,1}(X)$, the number of Kahler moduli.
\item[(C3)] The bundle $V$ needs to be equivariant under the discrete symmetry $\Gamma$ for it to descend to a bundle $\hat{V}$ on the quotient manifolds $\hat{X}=X/\Gamma$. In order to preserve the splitting type, we realize this by asking all $U_a$ to be $\Gamma$-equivariant individually. Hence, they descend to bundles $\hat{U}_a$, so that $\hat{V}=\oplus_{a=1}^f\hat{U}_a$.  In such cases, we can characterize the $U_a$ equivariant structures by $\Gamma$-representations $\chi_a^*$.  
\end{itemize}

\subsubsection{Conditions on spectrum of GUT theory}
\label{sec:condgutspec}

In preparation for considering specific examples, we list here a set of conditions that we must impose on split heterotic models that could lead to standard model-like physics in the associated four-dimensional theory. Some of the physical conditions on the spectrum can already be formulated in terms of the underlying GUT theory and are, therefore, somewhat easier to check.
\begin{itemize}
\item[(S1)] In order to have a chiral asymmetry of three families we require that
\begin{equation}
 -{\rm ind}(V)=\sum_aN({\bf 10}_{{\bf e}_a})\stackrel{!}{=}3|\Gamma|\; . \label{S1}
\end{equation}
Since the index divides by the group order, $|\Gamma|$, as we descend to the downstairs bundle $\hat{V}$ this guarantees a chiral asymmetry of three ${\bf 10}$ multiplets. The total chiral asymmetry of $\bar{\bf 5}$ multiplets equals $-{\rm ind}(\wedge^2 V)$. However, since ${\rm ind}(\wedge^2 V)={\rm ind}(V)$ for an $SU(5)$ bundle no additional condition is needed for the $\bar{\bf 5}$ multiplets.
\item[(S2)] We would like to exclude any mirror families originating from $\bar{\bf 10}$ multiplets. It is possible (although perhaps not likely) that an entire $\bar{\bf 10}$ multiplet is projected out by the Wilson line. Also, ${\bf 10}$-$\bar{\bf 10}$ pairs might be lifted away from the split locus, when singlet VEVs are switched on. However, we would like to use a cleaner approach and remove $\bar{\bf 10}$ multiplets completely at the GUT level and on the split locus. Hence, we require that
\begin{equation}
 \sum_a n(\bar{\bf 10}_{-{\bf e}_a})=h^2(X,V)\stackrel{!}{=}0\; .
\end{equation} 
\item[(S3)] In the downstairs spectrum we would like to have at least one pair of Higgs doublets. A necessary condition for this is the existence of at least one ${\bf 5}$--$\bar{\bf 5}$ pair in the GUT theory. This means that
\begin{equation}
 \sum_a n({\bf 5}_{-2{\bf e}_a})+\sum_{a<b}n({\bf 5}_{-{\bf e}_a-{\bf e}_b})=h^2(X,\wedge^2V)\stackrel{!}{>}0\; .
\end{equation} 
\end{itemize}

If all conditions above are satisfied we are guaranteed a model with the standard model gauge group (times $U(1)$ symmetries, possibly anomalous) and precisely three families of quarks and leptons (and no mirror families). The only additional multiplets are whatever remains from the ${\bf 5}$--$\bar{\bf 5}$ pairs after including the Wilson line. We need to ensure that the Higgs triplets are projected out and at least one pair of Higgs doublets survives. 
\begin{itemize}
\item[(H1)] In order to remove all Higgs triplets we demand that
\begin{equation}
n(\bar{T}_{a})=h^2(X,\wedge^2U_a,\chi_a\otimes\chi_a\otimes W)\stackrel{!}{=}0\; ,\quad
n(\bar{T}_{a,b})=h^2(X,U_a\otimes U_b,\chi_a\otimes\chi_b\otimes W)\stackrel{!}{=}0\; .
\end{equation}
As mentioned earlier, for an Abelian group $\Gamma$, a sufficient (although not strictly necessary) condition for the Higgs triplets to be projected out in a particular sector $\wedge^2U_a$ or $U_a\otimes U_b$ is
\begin{equation}
 h^2(X,\wedge^2U_a)\stackrel{!}{<}|\Gamma|\; ,\quad h^2(X,U_b\otimes U_c)\stackrel{!}{<}|\Gamma|\; .
\end{equation} 
\item[(H2)] In order to keep at least one pair of Higgs doublets we demand that
\begin{equation}
n(\bar{H}_{a})=h^2(X,\wedge^2U_a,\chi_a\otimes\chi_a\otimes\tilde{W})\stackrel{!}{>}0\; \quad
 n(\bar{H}_{a,b})=h^2(X,U_a\otimes U_b,\chi_a\otimes\chi_b\otimes\tilde{W})\stackrel{!}{>}0
\end{equation} 
in at least one sector $\wedge^2U_a$ or $U_a\otimes U_b$.
\end{itemize}
If these two further conditions are satisfied, we have a standard model spectrum, that is a standard model gauge group (times $U(1)$ factors), precisely three families of quarks and leptons, one or more pairs of Higgs doublets and no exotic fields charged under the standard model group.

\section{Heterotic line bundle models on CICYs}

In this section, we will begin a systematic and algorithmic search for split heterotic model with a standard model particle spectrum, that is, models which satisfy the constraints (C1)--(C3), (S1)--(S3) and (H1), (H2) described in the previous section. As our starting point, we will focus on ``maximally split" models defined by a sum of five line bundles
\beq\label{Vline}
V= \bigoplus_{a=1}^5 L_a\; .
\eeq
That is, we consider purely Abelian internal gauge fields such that all higher-rank factors $H_s$ in the split structure group given in \eref{HsJ} vanish. This choice is motivated by starting with the most basic splitting type and by the relative technical simplicity of line bundles. In particular, the slope-stability conditions of Section \ref{split_stab} reduce to the straightforward zero-slope condition of \eref{slopecond} in this case. Starting from line-bundle models we still have a window into a larger moduli space of geometries via the bundle deformations discussed in Section \ref{sec:dtermsgen}. Moreover, as described in Section \ref{sec:database}, it is frequently of phenomenological interest to turn on singlet VEVs, thereby continuing to other splitting types. 

In order to systematically construct a large-scale data set of line bundle models, we must specify an explicit class of Calabi-Yau geometries and the available line bundles over them. In this paper, we will take as our arena perhaps the simplest set of Calabi-Yau three-folds, namely complete intersections in products of projective spaces (CICYs), and line bundles defined over them. The set of CICYs have been completely classified \cite{Candelas:1987kf,Gagnon:1994ek}  and there exists a list~\cite{the_cy_list} of $7890$ such spaces together with their basic properties. 

In addition, the two main technical issues relevant for the construction of heterotic line bundle models have been resolved for CICYs. First, freely acting discrete symmetries of CICY manifolds have been classified recently in Ref.~\cite{Braun:2010vc}\footnote{This classification covers discrete groups which can be constructed via a linear action on the coordinates of the ambient projective spaces.}. Secondly, efficient methods to compute line bundle cohomology on CICYs have been described and implemented by the authors in Ref.~\cite{Anderson:2007nc,Anderson:2008uw,Anderson:2008ex}. Our constructions will heavily rely on both of these technical results. 

We begin by reviewing the relevant properties of CICYs and line bundles over them before we move on to the construction of models. Since the Picard number, that is, the dimension of the space of available line bundles, is given by the Hodge number $h^{1,1}(X)$ of the underlying three-fold, $X$, one expects the number of models and their complexity to increase with $h^{1,1}(X)$. We will, therefore, start with CICYs with the smallest possible Picard number $h^{1,1}(X)=2$ (recall that the vanishing slope constraint~\eqref{slopecond} cannot be satisfied for $h^{1,1}(X)=1$) and work our way up to and including $h^{1,1}(X)=5$ (the smallest Picard number for which models free of massless $U(1)$ vector fields can be constructed). Of course, on each of those CICYs we must study a large number of line bundle sums in order to find those with a phenomenologically interesting particle spectrum.

\subsection{Complete intersection Calabi-Yau three-folds}
Complete intersection Calabi-Yau manifolds $X$ are defined as the common zero locus of $K$ homogeneous polynomials on the ambient space $\cA =\IP^{N_1} \times \ldots \times \IP^{N_m}$, given by a product of $m$ projective spaces with dimensions $N_i$. To obtain three-folds from such complete intersections we obviously need $\sum_{i=1}^mN_i-K=3$. The defining polynomials, $p_\alpha$, are characterised by their multi-degrees ${\bf q}_j=(q_\alpha^1,\ldots , q_\alpha^m)$, where $q_\alpha^i$ specifies the degree of $p_\alpha$ in the coordinates of the factor $\IP^{N_i}$ in $\cA$. This information can be summarised by the configuration matrix
\beq\label{cy-config}
\left[\ba{c|cccc}
\IP^{N_1} & q_{1}^{1} & q_{2}^{1} & \ldots & q_{K}^{1} \\
\IP^{N_2} & q_{1}^{2} & q_{2}^{2} & \ldots & q_{K}^{2} \\
\vdots & \vdots & \vdots & \ddots & \vdots \\
\IP^{N_m} & q_{1}^{m} & q_{2}^{m} & \ldots & q_{K}^{m} \\
\ea\right]_{m \times K}\; .
\eeq
whose columns are the multi-degrees of the defining polynomials. In order to obtain a Calabi-Yau manifold, that is, a manifold with vanishing first Chern class, the conditions $\sum_{\alpha=1}^Kq^i_\alpha =N_i+1$ need to be satisfied for all $i$, so that each row of the configuration matrix sums up to the dimension of the associated projective space plus one. 

As stated above, the classification of CICYs has led to a list of 7890 such configuration matrices (although not all of them correspond to different Calabi-Yau three-folds~\cite{Anderson:2008uw}). For the purpose of this paper, we will only be interested in the low Picard number cases within this set, satisfying $2\leq h^{1,1}(X)\leq 5$, which, in addition have at least one symmetry of the form classified in \cite{Braun:2010vc}. There are $65$ such configurations although only $39$ of them turn out to be inequivalent\footnote{It is still useful valuable to have the same Calabi-Yau three-fold represented by different CICY configurations since the classification of discrete symmetries, which are restricted to act linearly on the ambient space, can depend on the representation used.}. These $65$ manifolds, together with their main properties are available at the website listed in Ref.~\cite{the_cy_list}.

We need to collect a few general properties of these manifolds which will be used in the following. 
It turns out that all of the $65$ manifolds relevant to our initial scan are ``favourable" in the sense that their complete second cohomology descends from the ambient space. Hence, if we introduce the standard Kahler forms $J_i$, normalised as $\int_{\mathbb{P}^{N_i}}J_i^{N_i}=1$, on the projective factors of the ambient space their restriction to $X$ (which we denote by the same symbol) spans the complete second cohomology of $X$. If we introduce the forms $\rho_X=\wedge_{\alpha=1}^K(\sum_iq^i_\alpha J_i)$ the triple intersection numbers in the basis $\{J_i\}$ can be easily computed from the formula
\begin{equation}
d_{ijk}=\int_XJ_i\wedge J_j\wedge J_k=\int_{\cal A}J_i\wedge J_j\wedge J_k\wedge\rho_X\; .
\end{equation}
Elements of the fourth cohomology (and, by Poincar\'e duality, the second homology) of $X$ will often be given relative to a basis $\{\nu^i\}$ of four forms dual to $\{J_i\}$, satisfying $\int_XJ_i\wedge \nu^j=\delta_i^j$.  
For example, we expand the second Chern class of $X$ as $c_2(TX)=c_{2i}(TX)\nu^i$ and the values of the coefficients $c_{2i}(TX)$ are explicitly given in the data available from the database \cite{database}. Kahler forms $J$ on $X$ can be expanded as $J=t^iJ_i$, where $t^i$ are the Kahler moduli. For favourable CICYs,  the Kahler cone is given by $t^i>0$ for all $i$ and effective classes in the second homology of $X$ corresponds to positive linear combinations of the four-forms $\nu^i$. 

\subsection{Line bundles on favourable CICYs}
Line bundles on Calabi-Yau three-folds are classified by their first Chern class (modulo discrete torsion which determines Wilson Lines) and, on favourable CICYs, they can therefore be labelled by an integer vector ${\bf k}=(k^1,\ldots, k^m)$. We denote by $L=\cO_X({\bf k})$ the line bundle with first Chern class $c_1(\cO_X({\bf k}))=k^iJ_i$. This line bundle can also be thought of as the restriction of the ambient space line bundle $\cO_{\cal A}({\bf k})=\cO_{\mathbb{P}^{N_1}}(k^1)\otimes\dots\otimes \cO_{\mathbb{P}^{N_m}}(k^m)$ to $X$. Its dual, $L^*$, is simply given by $L^*=\cO_X(-{\bf k})$. With the Todd class of a Calabi-Yau three fold given by ${\rm Td}(TX)=1+c_2(TX)/12$ it is straightforward to find the index
\begin{equation}\label{ind}
\ind(L)  \equiv\sum_{q=0}^{3}(-1)^{q}h^{q}(X,L) = \int_X\ch(L)\wedge\td(X)
 =\frac{1}{6}\left( d_{ijl} k^ik^jk^l+\frac{1}{2}k^ic_{2i}
(TX)\right)
\end{equation}
of $L$. However, computing the line bundle cohomology groups $H^q(X,L)$ individually is more involved. One useful relation is provided by Serre duality \cite{AG} which, for Calabi-Yau three-folds, states that $H^q(X,L)\cong H^{3-q}(X,L^*)$. 
Positive (ample) line bundles in the present context are those for which all $k^i>0$. To such positive line bundles we can apply Kodaira's vanishing theorem \cite{AG} which states that $H^q(X,L)=0$ for $q>0$. Hence, for positive line bundles the zeroth cohomology is the only non-vanishing one which can be computed from the above index formula alone. Similarly, negative line bundles are those for which all $k^i<0$. Their only non-vanishing cohomology is the third which again can be computed from the index. Unfortunately, positive and negative line bundles are not useful for our model building purposes. In fact, for such line bundles the zero slope condition \eqref{slopecond} cannot be satisfied, given that the Kahler cone for our three-folds is given by $t^i>0$. Hence, we need to understand the cohomology of ``mixed" line bundles. In this context, there exist weaker vanishing theorems which can sometimes be helpful. For example, the vanishing condition mentioned in footnote \ref{footnote5}, which states that if the slope $\mu(L)=d_{ijk}c_1^i(L)t^kt^k$ is negative somewhere in the Kahler cone, then $H^0(X,L)=0$ \cite{kob}, can be useful in determining the cohomology of many line bundles of interest to us.

To extract further information about line bundle cohomology one can consider Koszul resolutions combined with Bott-Borel-Weil representations \cite{Hubsch:1992nu} of ambient space cohomology, as described in Ref.~\cite{Anderson:2008ex}. These methods, combined with the above vanishing theorems, allow us to calculate the vast majority of line bundle cohomology groups and underlie most of the cohomology results in this paper. For a small number of line bundles these methods do not lead to a complete answer and one has to resort to more explicit methods, such as computing Cech cohomology \cite{AG}. The method for computing graded line bundle cohomology on CICYs is outlined in Appendix~\ref{appB}.\\

In our systematic scans for heterotic line bundle standard models on CICYs, we will construct rank five bundles $V$ with structure group $S(U(1)^5)$ by considering sums~\eqref{Vline} of five line bundle $L_a=\cO_X(k_a^i)$, satisfying $c_1(V)=\sum_ac_1(L_a)=0$. This corresponds to the splitting type ${\bf n}=(1,1,1,1,1)$ with $f=5$ from the general set-up in Section~\ref{genform} and all basic results from this section, in particular the content of Tables~\ref{tab2} and \ref{tab:gc}, apply with the identification $U_a=L_a$. In addition, one special feature of line bundle models has to be taken into account, namely the vanishing of the bundles $\wedge^2 U_a$,  $\wedge^2 U_a^*$ and, hence, the absence of the corresponding $\bar{\bf 5}_{2{\bf e}_a}$, ${\bf 5}_{-2{\bf e}_a}$ multiplets given in Table~\ref{tab2}. For the Chern characters and the index of such a sum of line bundles one has
\begin{eqnarray}
 {\rm ch}_1(V)&=&\sum_ak_a^i\stackrel{!}{=}0\\
 {\rm ch}_2(V)&=&\frac{1}{2}d_{ijl}\sum_a k_a^jk_a^l\label{ch2V}\\
 {\rm ind}(V)&=&\frac{1}{6}d_{ijl}\sum_ak_a^ik_a^jk_a^l\; .
\end{eqnarray} 
\subsection{The scan}
With the methods described above, we were able to perform the scan described in previous sections, over approximately $10^{12}$ different line bundle sums $k_a^i$, to build line bundle standard models over CICYs. The GUT models from this scan have already been presented and discussed in Ref.~\cite{Anderson:2011ns}. They were obtained by considering all favourable CICYs with $h^{1,1}(X)\leq 5$ and with freely-acting symmetries. No phenomenologically viable model were found for $h^{1,1}(X)=2,3$. The scan, which ranged over $-3\leq k_a^i\leq 3$ for $h^{1,1}(X)=4$ and $-2\leq k_a^i\leq 2$ for $h^{1,1}(X)=5$, led to 202 GUT models on $13$ CICYs. Here, we have calculated the explicit downstairs models with standard model gauge group from all these GUT models and the results are presented in the database~\cite{database}. Phenomenological properties of these models have already been discussed in Section~\ref{sec:database}. 

\section{Further consequences of split Heterotic geometry}\label{further_cons}
In this section we explore the possibility of more complicated equivariant structures on the bundle $V$, such that not every constituent bundle $U_a$ is equivariant by itself. Such models are relatively rare at least among line bundle models. For example, the scan in Ref.~\cite{Anderson:2011ns} which led to $202$ phenomenologically promising GUT models, only produced four models with non-trivial equivariant building blocks. Nevertheless, such constructions might be of interest for a number of phenomenological reasons, in particular in relation to the up-type Yukawa matrix and the possibility of a Peccei-Quinn symmetry. We begin by outlining the theoretical issues which arise in these constructions and discuss possible phenomenological applications towards the end of the section.

\subsection{Line bundle models with non-trivial equivariant blocks}\label{nontrivial_equiv}
In this section, we explore sums of line bundles $V=\bigoplus_a L_a$, for which not all of the $L_a$ admit an equivariant structure. Such sums can still descend to the quotient manifold, ${\hat X}$, if $V$ splits into equivariant blocks, that is, into non-trivial line bundle sums which are equivariant as a whole even thought their constituent line bundles are not. A simple example of this type of non-trivial structure can be found by considering the line bundle $\cO(1)$ on the quintic hypersurface in $\mathbb{P}^4$. This manifold admits a well-known $\Gamma=\mathbb{Z}_5 \times \mathbb{Z}_5$ freely acting discrete symmetry. On the homogenous coordinates of $\mathbb{P}^4$, $\Gamma$ can be represented as
\beq\label{quin_gamma}
\mathbb{Z}_{5}^{1}:~~x_k \to x_{k+1}~~,~~\mathbb{Z}_{5}^{2}:~~x_{k} \to \alpha^k x_k
\eeq
where $\alpha$ is a fifth root of unity, so $\alpha^5=1$. It is straightforward to verify that $\cO(1)$ is not equivariant with respect to the symmetry \eref{quin_gamma}. Indeed, not only is ${\rm ind}(\cO(1))=5$ which is clearly not divisible by $|\Gamma|=25$, but it is straightforward to show that, for the induced action of \eref{quin_gamma} on $H^0(X, \cO(1))$, the two $\mathbb{Z}_5$ symmetries do not commute and instead form a representation of the order $125$ Heisenberg group \cite{Donagi:2003tb} (see Appendix~\ref{appA} for mathematical details on equivariance). However, a sum of five such line bundles
\beq\label{vquin}
V=\cO(1)^{\oplus 5}
\eeq
does admit an equivariant structure. The required morphisms $\phi \in {\rm Hom}(V,V)$ are no longer a simple group character (as was the case for individually equivariant line bundles), but instead are generated by the following two matrices acting fiber-wise on the sum in \eref{vquin}
\beq\label{quin_equiv}
\phi_1=\left(
\begin{array}
[c]{ccccc}%
0 & 1&0&0&0 \\
0 & 0&1&0&0\\
0&0&0&1&0\\
0&0&0&0&1\\
1&0&0&0&0 
\end{array}
\right),~~\phi_2=\left(
\begin{array}
[c]{ccccc}%
1 & 0&0&0&0 \\
0 & \alpha^2&0&0&0\\
0&0&\alpha^4 &0&0\\
0&0&0&\alpha^1&0\\
0&0&0&0&\alpha^3
\end{array}
\right) \;.
\eeq
It can be verified that under the combined action~\eqref{sactionapp} of \eref{quin_gamma} and \eref{quin_equiv} the sections $H^0(X, \cO(1)^{\oplus 5})$ carry a representation of $\Gamma=\mathbb{Z}_5 \times \mathbb{Z}_5$.

In this example, it is clear that $V$ in \eref{vquin} is indeed an equivariant bundle and equal to the pull-back of some rank five bundle, $\hat{V}$ on $\hat{X}=X/\Gamma$. However, in this case the ``downstairs" bundle, $\hat{V}$, is {\it not a sum of line bundles}, rather is some generically indecomposable rank five vector bundle on $\hat{X}$. Since $H^0(X,\hat{V}^*\otimes \hat{V})=1$, ${\hat V}$ is in fact a simple bundle (in the algebro-geometric sense) \cite{AG}. In other words, given the covering map $q: X \to X/\Gamma$, the operation of pulling back, $q^*(\hat{V})$, need not preserve the structure group. In particular, a non-split bundle may pull-back to a split one.

Such bundles are of interest to us in the present context. Although they represent more complicated, higher-rank objects over ${\hat X}$, their properties are completely determined by a sum of line bundles over $X$ and, hence, are easily analyzed using the techniques we have described in this work. Although we expect relative few models with non-trivial equivariant building blocks they may have several features of phenomenological interest, as we will see. Before we begin to explore these features in detail, however, it should be noted that for such models the number of Green-Schwarz anomalous $U(1)$ symmetries will be different in the ``upstairs" and ``downstairs" theories. For example, consider a sum of two line bundles $L_1 \oplus L_2 \subset V=\bigoplus_{a=1}^5 L_a$ which are equivariant as a pair but not individually. If $c_1(V)=0$ there will be four $U(1)$ symmetries upstairs, but only three associated  to the downstairs bundle, which has the split form
\beq
\hat{V}=U_1 \oplus L_3 \oplus  L_4 \oplus  L_5
\eeq
with $U_1$ an indecomposable, $U(2)$ bundle. In the extreme case where all five line bundles form an equivariant block (as in the above toy example on the quintic), there would in fact be no enhanced $U(1)$ symmetries left in the downstairs theory. It seems in such cases one has lost not only all additional $U(1)$ symmetries but also the constraints they normally imply for the structure of the four-dimensional theory. However, since the downstairs theory associated to $(\hat{X}, \hat{V})$ is derived as the $\Gamma$-invariant part of the theory on $(X, V)$, the $U(1)$ symmetries which appear in the upstairs theory can still constrain the theory. For example, the downstairs Yukawa couplings $\lambda^{(u)}_{pqr}$ for ${\bf 10}^p{\bf 10}^q{\bf 5}^r$ (see Eq.~\eqref{WYuk}) are determined by the Yoneda pairing
\beq\label{yoneda_down}
Y_{\rm downstairs}=\int_{\hat{X}} H^1(\hat{X}, \hat{V})\wedge H^1(\hat{X}, \hat{V}) \wedge H^1(\hat{X},\wedge^2\hat{V}^*)\; .
\eeq
This can be written in terms of the invariant part of the upstairs Yukawa couplings as
\beq\label{yoneda_up}
\text{Invariant}_{\Gamma,\phi} \left( \int_{X} H^1(X, q^*(\hat{V}))\wedge H^1(X, q^*(\hat{V}))\wedge H^1(X, q^*(\wedge^2 \hat{V}^*)) \right)\; .
\eeq
While $H^1(X, q^*(\hat{V}))$ need not be the cohomology of a simple line bundle $L$ on $X$, the integral in \eref{yoneda_up} will in general be constrained to be $U(1)$ gauge invariant for all the anomalous $U(1)$ symmetries which arise in the upstairs theory. With this observation in hand we turn now to look at some of the phenomenological features which arise in models with non-trivial equivariant blocks. 

\subsection{Up-type Yukawa couplings}
As discussed in Section~\ref{sec:lbsms} (see Eq.~\eqref{WYuk}), for line bundle models with $V=\bigoplus_a L_a$, the up-type Yukawa couplings 
\begin{equation}
 \lambda^{(u)}_{pq}\bar{H}{\bf 10}^p{\bf 10}^q \label{upyuk}
\end{equation}
are allowed only if the Higgs charge $-{\bf Q}(\bar{H)}={\bf e}_{\bar h}+{\bf e}_{\bar g}$ equals ${\bf Q}({\bf 10}^p)+{\bf Q}({\bf 10}^q)={\bf e}_{a_p}+{\bf e}_{b_q}$. For pure line bundle models, we always have ${\bar h}<{\bar g}$ since the sector $H^1(X,\wedge^2 U_a)$ which may lead to up-type Higgs doublets with charge $-2{\bf e}_a$ is absent for line bundle models. Since the Yukawa terms in~\eqref{upyuk} are symmetric in family space it follows that up Yukawa couplings for such models without equivariant blocks cannot have rank one. This conclusion can obviously change away from the Abelian locus when singlet VEVs are switched on or when multiple Higgs pairs are considered. 

Another possibility to generate phenomenologically desirable rank one up Yukawa matrices at the perturbative level is to consider line bundle models with non-trivial equivariant blocks. For example, consider the case where two line bundles, $L_1 \oplus L_2$ are associated to the pull-back of a non-Abelian rank two bundle, ${\hat U}$ on the quotient space, ${\hat X}$. Then the downstairs Yukawa couplings may contain a terms of the form
\beq
Y_{\rm downstairs} \sim \int_{\hat{X}} H^1({\hat X},{\hat U})_{{\bf e}_a} \wedge H^1({\hat X}, {\hat U})_{{\bf e}_a} \wedge H^1({\hat X}, \wedge^2\hat{U}^*)_{-2{\bf e}_{a}}
\eeq
which can lead to a rank one mass matrix. Such a term is possible since each of the downstairs cohomology groups above, pulls back not to a single line bundle cohomology on $X$ but a sum of two line bundles. That is, $Y^{(u)}$ is given by the invariant part of
\beq
\int_{X} \left( \begin{array}{cc}
H^1(X, L_1)&  \\
 & H^1(X, L_2)\\
\end{array}
\right) \wedge
\left(
\begin{array}{cc}
H^1(X, L_1)&  \\
 & H^1(X, L_2)\\
\end{array}
\right) \wedge
H^1(X, L_{1}^{*} \otimes L_2^*)
\eeq
where the $2 \times 2$ matrices schematically represent the group invariant elements of cohomology that are obtained from non-trivial combinations of $H^1(X, L_1)$ and $H^1(X, L_2)$. 

\subsection{Realizing a Peccei-Quinn symmetry}

Peccei-Quinn $U(1)$ symmetries, denoted by $U_{\rm PQ}(1)$, may be used to explain some of the more puzzling aspects of supersymmetric GUTs. Of particular interest is their role in addressing the $\mu$ problem, that is, the problem of why the superpotential Higgs mass term is so small given that it is a super-renormalizable operator. The class of models explored in this work, perhaps point at least in the right direction since the Higgs doublets are precisely massless at the split locus of the bundle. However this feature is not guaranteed to persist away from the split locus when non-trivial singlet VEVs are switched on and moduli stabilization might well drive some models to such a region in moduli space. However, if the $\mu$ term is forbidden by a $U_{\rm PQ}(1)$ symmetry it is more robustly protected. Peccei-Quinn symmetries can also be motivated from proton stability. Assuming neutrality of the Yukawa-couplings it is the easy to show that any $U(1)$ symmetry which forbids the dimension five proton decay operators $\bar{\bf 5}^p {\bf 10}^q {\bf 10}^r {\bf 10}^s$ (see Eq.~\eqref{Wprot5}) must in fact be a $U_{\rm PQ}(1)$ symmetry. For these reasons, it is important to study whether such a symmetry can be realized in split heterotic models \footnote{For analyses of similar issues in an F-theory context see for example \cite{Dudas:2010zb,Marsano:2010sq,Dolan:2011aq}.}.

We will now argue that in split models with bundle $V=\bigoplus_a U_a$ and each $U_a$ equivariant individually, no $U_{\rm PQ}(1)$ can arise. For such models, each constituent bundle $U_a$ descends to a bundle $\hat{U}_a$ on the quotient space $\hat{X}=X/\Gamma$ and it follows that ${\rm ind}(\hat{U}_a\otimes\hat{U}_b)={\rm ind}(U_a\otimes U_b)/|\Gamma|$ (and similarly for $\wedge^2U_a$), that is the index is divisible by the group order $|\Gamma|$ in each sector of $\bar{\bf 5}_{{\bf e}_a+{\bf e}_b}$ and ${\bf 5}_{-{\bf e}_a-{\bf e}_b}$ multiplets. Of course, the chiral asymmetry in each such sector must be zero or negative, ${\rm ind}(\hat{U}_a\otimes\hat{U}_b)\leq 0$, or else this sector leads to complete ${\bf 5}$ multiplets downstairs which contain phenomenologically unacceptable Higgs triplets. The $\bar{\bf 5}$ excess in each sector will contribute towards the families, while divisibility of the index implies that any remainders from the $\bar{\bf 5}$--${\bf 5}$ pairs have to come in vector-like pairs. In conclusion, if all constituent bundles $U_a$ are equivariant individually and we require Higgs triplets to be projected out by Wilson-line breaking, then the Higgs doublets can always be grouped into vector-like pairs under the $U(1)$ symmetries.

We would now like to argue, for the case of line bundle sums $V=\bigoplus_a L_a$, that this statement does not necessarily hold any more in the presence of non-trivial equivariant blocks. Suppose that the line bundle sum $L_1\oplus L_2\subset V$ constitutes a non-trivial equivariant block and is the pull-back of a non-decomposable rank two bundle $U$ on $\hat{X}$, that is $L_1\oplus L_2 =q^{*}({\hat U})$. Now consider a third line bundle, $L_a$ in $V$ such that ${\rm ind}((L_1 \oplus L_2)\otimes L_a)<0$ and $h^2((L_1 \oplus L_2)\otimes L_a)<|\Gamma|$. The Higgs multiplets which arise from $(L_1 \oplus L_2)\otimes L_a$ can be obtained from the up-stairs cohomology by extracting the part which corresponds to the appropriate $\Gamma$-representation, along the lines of Section~\ref{sec:symm}. If this process leads to Higgs multiplets 
\begin{equation}
 H\in H^1(X, L_{2}\otimes L_a)\; ,\quad \bar{H}\in H^1(X,L_{1}^{*}\otimes L_a^*)\; ,
\end{equation}
a pattern which presumably can be arranged for appropriate model building choices, then it is clear that a linear combination of the upstairs $U(1)$ symmetries will behave as a $U_{\rm PQ}(1)$ symmetry and forbid the $\mu$-term as well as dangerous dimension five proton decay operators. It is worth noting that in this case, although the $\mu$-terms of the final four-dimensional theory are effectively forbidden by the $U(1)$ invariance described above, there is no corresponding $U(1)$ gauge symmetry in the four-dimensional theory. \\

It should be clear from the above discussion that many features of the four-dimensional theory can be affected by non-trivial equivariant blocks, often in subtle and surprising ways. We hope to explore the consequences of such behaviour in more detail in future work. For now, it should be noted that, unfortunately, the interesting features described above do not arise in the four models with non-trivial equivariant blocks in our current data set. All four models suffer from the problem that the color triplets cannot be projected out. If desired, one could modify the scanning criteria we have employed to generate models with a $U_{\rm PQ}(1)$ using the methods described in this paper.

\section{Summary}
\label{sec:summary}
In this paper, we have developed the formalism for heterotic models with split vector bundles, with particular emphasis on the maximal splitting into line bundle sums. We have shown that such heterotic line bundle models are a promising arena for building particle physics models within string theory and we have presented a database~\cite{database} of about $400$ heterotic line bundle standard models. All of these models have a standard model gauge group times four additional and frequently Green-Schwarz anomalous $U(1)$ symmetries, the exact matter spectrum of the MSSM, one or more pairs of Higgs doublets, a spectrum of bundle moduli which are standard model singlets and no exotics charged under the standard model group of any kind. 

The additional $U(1)$ symmetries constrain the allowed operators in the four-dimensional theory and provide an interesting tool for low-energy phenomenology. For our line bundle standard models we have worked out the spectrum of $U(1)$ invariant four-dimensional operators, including operators with insertions of the bundle moduli singlets. The results which are presented in the database~\cite{database} can be used to discuss a large variety of phenomenological issues, including the structure of Yukawa couplings, proton stability, the $\mu$-term and neutrino masses. In Section~\ref{sec:database} these issues have been illustrated for a particular example model from the database. We have emphasized that this specific model has not been put forward as the most attractive one, but merely as an example to demonstrate the phenomenological possibilities of heterotic line bundle models. We hope that the database~\cite{database} provides a starting point for exploring these phenomenological questions in depth and, eventually, to isolate a true string standard model which reproduces all known features of low-energy physics.

The present work is merely the first step in this direction. First and foremost we need to extend the current scans for line bundle standard models. The present models, collected in the database~\cite{database} are defined on a relatively small set of Calabi-Yau three-folds and rely on line bundles with a fairly restricted range of integer entries. Both of these restrictions can be lifted, at least to a certain extent, and we expect to find many more line bundle standard models in this case. Ideally, this should eventually  lead to a classification of all line bundle standard models on the known sets of Calabi-Yau three-folds. Such a data set, which will likely contain tens of thousands or possibly even more models, would be a legitimate starting point for the search for a true string standard model.

\subsubsection*{Acknowledgments}
L.A. is supported by the Fundamental Laws Initiative of the Center for the Fundamental Laws of Nature, Harvard University. J.~G. would like to acknowledge support by the NSF-Microsoft grant NSF/CCF-1048082. A.~L.~is supported in part by the EC 6th Framework Programme MRTN-CT-2004-503369 and EPSRC network grant EP/I02784X/1. He would like to thank the string theory group at LMU Munich for hospitality. The work of EP is supported by a Marie Curie Intra European Fellowship within the 7th European Community Framework Programme. 

\appendix
\section{Equivariant Structures and Wilson Lines}
\label{appA}

One of the requirements we impose on line bundles sums
$U$ is that it respects the freely-acting symmetry, $\Gamma$, by which we quotient our
Calabi-Yau three-folds. More precisely, we need that our bundle $U
\stackrel{\pi}{\longrightarrow} X$ descends to a bundle $\hat{U}$ on
$X/\Gamma$, in the sense that $ U \cong q^* \hat{U}$. For a bundle to
descend to the quotient space it is necessary that the automorphisms
$\Gamma$ of $X$ ``lift'' to automorphisms of the bundle $U$ over $X$. In
other words, for each $g \in G$ there must exist a bundle morphism,
$\phi_g: U \to U$ which commutes with the projection $ \pi: U \to X$
and covers the action $g:X \to X$ on the base. Such a lifting of the
group action is known as an invariant structure on $U$. We can express
this concisely by the commutativity of the following diagram for all
$g \in \Gamma$.
\begin{equation}
  \begin{array}{lllll}
  &U&\stackrel{\phi_g}{\longrightarrow}&U&\\
  \pi &\downarrow&&\downarrow&\pi\\
  &X&\stackrel{g}{\longrightarrow}&X&
 \end{array}
 \label{invaltapp}
\end{equation} 
In addition to invariance we must require that the $\phi_g$ satisfy
what is called the co-cycle condition, namely that $\forall \; g,h \in
\Gamma$,
\bea \phi_g \circ \phi_h = \phi_{gh} \;.  \label{apppy1} \eea An invariant structure
which satisfies the cocycle condition is called an equivariant
structure on $U$.  If $U$ admits such a set of morphisms it is said to
admit an equivariant structure and, in this case, it descends to a
bundle $\hat{U}$ on $X/\Gamma$. Indeed, the set of vector bundles on $X/\Gamma$
is in one-to one correspondence with the set of equivariant vector
bundles on $X$.

How does one decide whether a sum of line bundles admits such an
equivariant structure? The bundle morphisms $\phi_g$ induce linear maps $\Phi_{g}$ on the global sections, $s\in H^0(X,U)$, of $U$ which are defined by
\begin{equation}
 s\rightarrow \Phi_{g}(s)=\phi_{g}\circ s\circ g^{-1}\; . \label{sactionapp}
\end{equation}
From Eq.~\eqref{apppy1}, these induced actions on the sections satisfy
\begin{equation}
\Phi_{g}\circ\Phi_{h}=\Phi_{gh}\; , \label{phitreprapp}
\end{equation}
that is, they form a $\Gamma$ representation on the sections of $U$. If the bundle $U$ is globally generated by its sections this provides a practical way of checking equivariance (see Ref.~\cite{Anderson:2009mh} for a more detailed discussion of this point). In such a case one can reconstruct the bundle morphisms $\phi_g$ from the action on section $\Phi_g$ and so $U$ is equivariant if and only if such linear maps on the sections exist. Since the sections of $U$ are usually given by polynomials in the ambient space homogeneous coordinates, checking for the existence of such maps on the sections is a practical possibility. If the bundle $U$ is not globally generated, as will frequently be the case for our examples, this method does not immediately apply. However, in this case, one can consider the twisted bundle $U\otimes L^p$, where $L$ is an ample line bundle known to be equivariant and $p>0$ is a sufficiently large integer such that the twisted bundle is globally generated. Equivariance of $U\otimes L^p$ can then be checked on its sections as described above and it is equivariant if and only if the original bundle $U$ is.  

To consider the actions of symmetries on sections, we shall start by considering symmetry actions on homogeneous polynomials. The two cases of interest to us here are where $\Gamma$ is a single $\mathbb{Z}_m$ factor, or a direct product of two such groups. In the discussion to follow much of the complication which will occur is only relevant for the case where we have more than one Abelian factor, and even then only for line bundles which must appear with a non-trivial multiplicity if they are to admit an equivariant structure. Here, we will discuss the general case as, once this is understood, specializing to the more straightforward examples is trivial. In particular it is very easy to see, by simplifying the following analysis, that for $\Gamma=\mathbb{Z}_m$ all line bundles admit an equivariant structure.

Consider a group $\Gamma=\mathbb{Z}_{m_1}
\times \mathbb{Z}_{m_2}$ acting on the ambient space ${\cal A}=\mathbb{P}^{N_1} \times \ldots
\times \mathbb{P}^{N_m}$. We will use indices $i,j,\ldots$ to label
projective space factors and $r,s, \ldots$ to label the Abelian
factors in $\Gamma$. For such an action to be well defined on the product
of projective spaces we have,
\begin{equation} M_r^{m_r} \left( \ba{c} a_0^i \\ $\vdots$ \\ a_{N_i}^i \ea \right)
= \lambda^i_r \left( \ba{c} a_0^i \\ $\vdots$ \\ a_{N_i}^i \ea \right)
\;\; \forall a,i \;, \end{equation}
where the $a^i$ are the homogeneous coordinates on the $i$'th projective space and $M_r$ is the action of the $\mathbb{Z}_{m_r}$ factor in $\Gamma$. In addition, we have the following to ensure that the Abelian factors commute when considered as an action on the projective spaces.
\be \label{abelian1} M_r M_s \left( \ba{c} a_0^i \\ $\vdots$ \\ a_{N_i}^i \ea \right) =
\Gamma^i_{rs} M_s M_r \left( \ba{c} a_0^i \\ $\vdots$ \\ a_{N_i}^i \ea
\right) \ee
Note that by definition we have $\Gamma_{sr}^i =(
\Gamma^i_{rs})^{-1}$. Given such an action on the homogeneous
coordinates one can compute the action on a vector of polynomials of multi-degree
$d^{\alpha}=\left[ d_1^{\alpha}, \ldots, d^{\alpha}_m \right]$, which
we denote $P_{d^{\alpha}}$ (note here that $\alpha$ labels the polynomial and $i=1,\ldots,m$ the projective spaces)\footnote{Strictly, we should use a different symbol for the action on vectors of polynomials and the homogenous coordinates. In this appendix, however, we will use $M_a$ for both in order to induce a less cluttered notation. Which action is meant is unambiguous from context.}.
\bea \label{sectionsapp}
M_r M_s P_{d^{\alpha}} &=& \gamma^{\alpha}_{rs} M_s M_r P_{d^{\alpha}} \\ \label{sections2app}
M_r^{m_r} P_{d^{\alpha}} &=& \Lambda^{\alpha}_r P_{d^{\alpha}} \\
\gamma_{rs}^{\alpha} &=& \Pi_{i=1}^m \left( \Gamma^i_{rs} \right)^{d^{\alpha}_i} \\ \label{sectionsapplast}
\Lambda^{\alpha}_r &=& \Pi_{i=1}^m \left( \lambda^i_r
\right)^{d^{\alpha}_i} 
\eea

Now that we have understood the action of our symmetries on vectors of polynomials
we can proceed to consider such polynomials to be the sections of globally generated sums line bundles on the products of projective spaces. Using the global generation property to induce the group action on the fibres from those of the sections we see that, in general, the naive induced action given in \eqref{sectionsapp} and \eqref{sections2app} need not give rise to an equivariant structure on the bundle. Equivalently, because of the possibility of projective rescalings, equations \eqref{sectionsapp} and \eqref{sections2app} do not obey a cocycle condition such as \eqref{phitreprapp} when considered as an action on sections. In order to ``fix up'' these transformations we include an additional bundle morphism $\hat{M}_r$ for each group action, which acts as a $GL(N)$ transformation on the sum of $N$ line bundles\footnote{To make a connection to the notation in our general discussion, the $M_r$ actions correspond to the $g^{-1}$ factor in \eqref{sactionapp} and the $\hat{M}_r$ correspond to the bundle morphisms $\phi_g$ which we must choose such that a cocycle condition of the form \eqref{phitreprapp} is satisfied. }. Note that in the case of a single $\mathbb{Z}_m$ we would only have equation \eqref{sections2app}, and as such we could always lay down an equivariant structure on any single line bundle, simply by taking the associated bundle morphism to be $\Lambda_r^{1/m_r}$ multiplied by the identity map.

Returning to the case of two Abelian factors in $\Gamma$, we then find the following requirements on the bundle morphisms from \eqref{phitreprapp}.
\bea
( \hat{M}_r M_r)^{m_r} P_d &=& P_d \\
&\Rightarrow& \hat{M}_r^{m_r} = \left( \ba{ccc} \Lambda^1_r & & \\ & \Lambda^2_r & \\ & & \ddots \ea \right) \; \; \textnormal{and} \\
(\hat{M}_r M_r) (\hat{M}_s M_s) P_d &=& (\hat{M}_s M_s) (\hat{M}_r M_r) P_d \\
&\Rightarrow& \hat{M}_r \hat{M}_s \left( \ba{ccc} \gamma^1_{rs} & & \\
  & \gamma_{rs}^2 & \\ & & \ddots \ea \right) M_s M_r P_d = \hat{M}_s
\hat{M}_r M_s M_r P_d 
\eea 
Thus demanding a good group action on the sum of line bundles on the ambient space, we obtain the conditions on our bundle morphisms $\hat{M}_r$, determined purely in terms of the original action on the homogenous coordinates, which follow.
\begin{equation} \label{hedgehog}
\hat{M}_r^{m_r}  \left( \ba{ccc} \Lambda^1_r & & \\ & \Lambda^2_r & \\ & & \ddots \ea \right) = 1 \;\; \forall \; a \; ,\quad
\hat{M_r} \hat{M_s} \left( \ba{ccc} \gamma^1_{rs} & & \\
  & \gamma_{rs}^2 & \\ & & \ddots \ea \right) = \hat{M}_s \hat{M}_r \;\; \forall \; r \neq s
\end{equation}  

In this paper we restrict ourselves to the case of individually equivariant line bundles or non-trivial equivariant blocks composed from a sum of the same line bundle~\footnote{In fact for the situation at hand, due to the structure of zeroth cohomologies of line bundles on projective spaces,  this simplification is true in generality. It is easy to show that, if one has a morphism $\phi_g$ which mixes different line bundles, its inverse $\phi_{g^{-1}}$ can not exist as one of the required homomorphism groups between line bundles will vanish. This of course means that we can not construct an equivariant structure on such a mixed sum of line bundles.}. Specializing to one such sum of identical line bundles for simplicity, we can set $d^{\alpha} = d^{\beta} \;\; \forall \; \alpha, \beta$. This simplification allows us to write $\Lambda^{\alpha}_r \cong \Lambda_r$ and
$\gamma^{\alpha}_{rs} \cong \gamma_{rs} \;\; \forall \;
\alpha$. Equations \eqref{hedgehog} then simplify to,
\bea \label{specialization}
\hat{M}_r^{m_r} \Lambda_r = 1 \;\; \textnormal{and} \;\;
\hat{M}_r \hat{M}_s \gamma_{rs} = \hat{M}_s \hat{M}_r
\eea
It is easy to show that if the equations \eqref{specialization} are to
have a solution then $\gamma_{rs}^{\hat{l}} =1$ where $\hat{l} =
\textnormal{LCM}(m_r, m_s)$. We can, without loss of generality,
choose $\hat{M}_1$ to be diagonal. Combining these two observations, it is
then possible to show that a solution to equations
\eqref{specialization}, giving rise to a good group action, only
exists if there we have a sum of $\hat{m}$ identical line bundles,
where ${\hat m}$ is the minimal integer such that $\gamma^{{\hat m}}=1$ (where $\gamma=\gamma_{12}$). In such a case the solution to the system takes the following form
\begin{equation}
\begin{array}{lllllll} 
\hat{M}_1 &=& \left( \ba{ccc} \tilde{M}_1 & & \\ & \tilde{M_1}  & \\ & &
  \ddots \ea \right)  &\textnormal{where}& \tilde{M}_1 &=& \Lambda_1^{-\frac{1}{N_1}} \left( \ba{cccc} 1 & & & \\ & \gamma & & \\ & &
  \ddots & \\ & & & \gamma^{{\hat m}-1} \ea \right)\\
\noalign{\vskip 6pt}
 \hat{M}_2 &=& \left( \ba{ccc} \tilde{M_2} & & \\ & \tilde{M_2} & \\ & & \ddots \ea \right) &\textnormal{where}&
 \tilde{M}_2 &=& \left( \ba{ccccc} 0&X_1&0&0& \ldots \\ 0 &0&X_2&0&\ldots \\ 
 \vdots & \vdots &\vdots& \vdots & \vdots \\ X_{\hat m} & 0 & 0 & 0 & \ldots \ea \right)
 \end{array}\; ,
\end{equation}
with $\left( \Pi_{i=1}^{m_2} X_i \right)^{\frac{m_2}{{\hat m}}} = \Lambda_2^{-1}$.
The above discussion allows us to lay down an equivariant structure on any sum of globally generated line bundles on products of projective spaces which admits an equivariant structure. Non-globally generated line bundles can be dealt with using this formalism by employing the technique of twisting by an ample equivariant line bundle as discussed earlier in this section. Finally we should consider equivariant structures on the restriction of these line bundles to the Calabi-Yau three-fold, defined as a complete intersection in the ambient space ${\cal A}$. Fortunately, for our favourable manifolds, due to the structure of the Koszul sequence and thanks to the fact that the normal bundle to the Calabi-Yau itself admits an equivariant structure, equivariant line bundle sums on the three-fold are in one-to-one correspondence with their ambient space counterparts.

\section{Representation content of cohomology}
\label{appB}

Based on the method to find an explicit equivariant structure on $U$ discussed in the previous appendix, we must now explain how to extract the representation content for cohomologies of $U$. If a bundle admits a $\Gamma$-equivariant structure, then the cohomology groups $H^q(X, U)$ form $\Gamma$-representations. We can then define the cohomology groups $H^q(X,U,R)$, the sub-spaces of $H^q(X,U)$ which transform under the $\Gamma$-representation $R$, and their dimensions $h^q(X,U,R)={\rm dim}(H^q(X,U,R))$. As in Table~\ref{tab:gc}, the physical properties of our models can then be expressed in terms of these ``graded" cohomologies. 

How do we compute these cohomologies? First of all, we relate the bundle $U$ to its ambient space counterpart ${\cal U}$ by the Koszul resolution. For illustration purposes let us consider a co-dimension one Calabi-Yau manifolds $X$ so that the Koszul resolution is given by a short exact sequence
\begin{equation}
 0\rightarrow {\cal N}^*\otimes {\cal U}\rightarrow {\cal U}\rightarrow U\rightarrow 0\; , \label{kos}
\end{equation}
where ${\cal N}$ is a bundle whose restriction to $X$ is the normal bundle of the Calabi-Yau manifold.
The cohomologies of the ambient space bundles ${\cal U}$ and ${\cal N}^*\otimes {\cal U}$ (and of bundles $\wedge^p {\cal N}^*\otimes{\cal U}$ for higher co-dimensions) can be expressed in terms of polynomials, using the Bott-Borel-Weil representation of cohomology on projective spaces~\cite{Hubsch:1992nu}. On those polynomial representatives we can explicitly act with the symmetry transformations in $\Gamma$ and, together with the equivariant structure on $U$, determined as explained in Appendix~\ref{appA}, the $\Gamma$-characters $\chi_{p,q}$ of $H^q({\cal A},\wedge^p {\cal N}^*\otimes{\cal U})$ can be computed. 
These characters are related to the characters $\chi_U^q$ of $H^q(X,U)$ by the long exact sequence in cohomology, associated to the Koszul resolution~\eqref{kos}. As an example, if $H^1({\cal A},{\cal N}^*\otimes{\cal U})=H^2({\cal A},{\cal N}^*\otimes{\cal U})=0$, so that $H^1(X,U)\cong H^1({\cal A},{\cal U})$, then we have $\chi^1_U=\chi^{0,1}$. In general, the characters $\chi^q_U$ can be expressed in terms of the ambient space characters $\chi^{p,q}$ by an obvious generalization of spectral sequence methods to characters~\footnote{In general, to get complete results, one needs to calculate the characters of images or kernels of maps in the spectral sequence. However, given the knowledge of the total cohomology dimensions $h^q(X,U)$, and the relative simplicity of our examples after imposing the relevant physical constraints, this is frequently unnecessary.}. 

The multiplicity of representations can, in general, be computed from the scalar product between two characters $\chi$ and $\psi$ defined as
\begin{equation}
 (\chi,\psi)=\frac{1}{|\Gamma|}\sum_{g\in \Gamma}\chi(g)\bar{\psi}(g)\; .
\end{equation}
Under this scalar product, the characters $\chi_\alpha$ of the irreducible $\Gamma$-representations form an orthonormal system, that is, $(\chi_\alpha,\chi_\beta)=\delta_{\alpha\beta}$. This means that the multiplicity $n^q_\alpha$ of the $\alpha^{\rm th}$ irreducible representation in $H^q(X,U)$ can be extracted by
\begin{equation}
 n^q_\alpha=(\chi_\alpha,\chi_U^q)\; .
\end{equation} 
For Abelian groups $\Gamma$, the case considered in this paper, all irreducible representations are one-dimensional so that the graded cohomologies are simply given by these multiplicities, that is $h^q(X,U,R_\alpha)=n^q_\alpha$. 


\end{document}